\documentclass[preprint,11]{aastex}

\begin{document}

\shorttitle{Wide-field study of NGC 1399}
\shortauthors{Dirsch et al.}

\title{The Globular Cluster System of NGC 1399. I. A Wide-field Photometric Study
\altaffilmark{3} }

\author{B. Dirsch \altaffilmark{1}, T. Richtler \altaffilmark{1}, D. Geisler \altaffilmark{1},
J.C. Forte \altaffilmark{2}, L.P. Bassino \altaffilmark{2}, W.P. Gieren \altaffilmark{1}} 

\altaffiltext{1}{Universidad de Concepci\'on, Departamento de F\'{\i}sica,
		Casilla 160-C, Concepci\'on, Chile}
\altaffiltext{2}{Facultad de Ciencias Astron\'omicas y Geof\'{\i}sicas,
		Universidad Nacional de La Plata, Paseo del Bosque S/N, 1900-La Plata,
		Argentina}
\altaffiltext{3}{Based partly on observations collected at the European Southern Observatory,
Cerro Paranal, Chile; ESO program 66.B-0393.}
\email{bdirsch@cepheid.cfm.udec.cl}

\begin{abstract} 

We present a photometric investigation of the globular cluster population of
NGC 1399, the central galaxy in the Fornax cluster, in Washington C and Kron R
filters within a field of $36\arcmin\times36\arcmin$, corresponding to about
200$\times$200 kpc at the Fornax distance.  This is the largest area around
this galaxy ever studied with CCD photometry. The cluster system of NGC 1399 is
found to extend further than 100 kpc away from the galaxy. The color
distribution exhibits a pronounced bimodality. 
Within a radial distance of about 55 kpc,
the red clusters are more strongly concentrated to the center than the blue
clusters.  At larger radii, the surface density profiles of the clusters are
indistinguishable and match well the distribution of the galaxy light. 
Over the entire radial distance range,
the surface brightness profile of NGC 1399 can be very well fitted by a power law
with an exponent of -1.85 and a core radius of $3.3\arcsec$. No steepening of
the luminosity profile can be detected at large radii.  We suggest that the
power law profile of NGC 1399 results from the galaxy being embedded in a large
dark matter halo, which prevents the stellar density profile from steepening
outwards.  The cluster system contains $6450\pm700$ clusters and the specific
frequency is found to be $5.1\pm 1.2$ in the V band. While NGC 1399 shows a
pronounced color gradient the nearby comparison galaxy NGC 1404 does not show
such a gradient.  Using simple assumptions about the underlying population that
formed during the same star formation event as the globular clusters, we present
a model in which we use radially changing local specific frequencies for the
red and blue subpopulations to fit the observations. We find that within
7\arcmin \ the required specific frequency of the blue clusters alone is a
factor of approximately 3 larger than that of the red ones. Outside this radius
however, both populations have a same high local specific frequency of around
8 and 13 (blue and red clusters, respectively).

\end{abstract}


\section{Introduction}

One of the central and challenging topics in the context of the structure and
evolution of early-type galaxies is the investigation of their globular
clusters systems (GCSs), how they formed, and what they might tell us about the
formation of their host galaxies (for recent reviews on the subject, see 
Ashman \& Zepf \cite{ashman98}, Elmegreen \cite{elmegreen99},
van den Bergh \cite{vandenbergh00}, Harris \cite{harris01}).

It has now emerged from numerous studies (e.g. Kundu \& Whitmore
\cite{kundu01a}, Kundu \& Whitmore \cite{kundu01b}, Larsen et
al.\cite{larsen01}, Gebhardt \& Kissler-Patig \cite{gebhardt99}, 
Geisler et al. \cite{geisler96b}, Whitmore et al. \cite{whitmore95}, Zepf\,\&\,Ashman
\cite{zepf93}) that GCSs of early-type galaxies are by no means homogeneous
populations, but exhibit substructure in terms of metallicity, kinematics and
spatial distribution. The formation of massive clusters does not occur in
isolation, but probably even needs to be accompanied by strong star forming
activity in general. This is apparent in ongoing merger events (see the list in
Harris \cite{harris01}) but also in the case of ``normal'' spiral galaxies
(Larsen \& Richtler \cite{larsen99}, \cite{larsen00}). Therefore, the presence
of GCs must be indicative of a corresponding field population with similar ages
and metallicities and thus GCs might be considered as tracers of this field
population, which otherwise is unaccessible due to the low surface brightness
of galaxy halos and/or because it is mixed up with other populations.

It might then be possible to trace back the population composition of an
elliptical galaxy, the distinct metallicities and spatial distributions of the
subcomponents, and compare these properties to what is expected if, for
example, the elliptical galaxy is a result of a merger (or multiple merger) of
disk galaxies (Bekki et al. \cite{bekki02}).  However, the relation of a GCS
and its components to the field components can be complicated.  Even if the
efficiency of cluster formation per unit mass had a universal value, as argued
by McLaughlin (\cite{mclaughlin99}), it does not mean that one can infer
directly the luminosity of the underlying field population from the number of
globular clusters because generally the fate of the gas is unknown.

A promising target to gain better insight into the connection between the
relation of cluster subpopulations to the underlying field population is NGC
1399, the central galaxy of the Fornax cluster. It has long been known to
possess a large number of globular clusters (Dawe\,\&\,Dickens \cite{dawe76})
out to large radii (Hanes\,\&\,Harris \cite{hanes86}, Harris\,\&\,Hanes
\cite{harris87}). Photographic studies of NGC 1399 showed that its luminosity
profile cannot be described by a de Vaucouleurs profile but by a power-law
(Schombert \cite{schombert86}, Killeen \& Bicknell \cite{killeen88}, Caon et
al. \cite{caon94}). Whether NGC 1399 is a cD galaxy or not depends on the
applied definition, which in turn differs from author to author. For example,
Bridges et al. \cite{bridges91} classify it as a cD, while Schombert
\cite{schombert86} does not include it in his list of cD galaxies. The spatial
distribution of galaxy light and globular clusters have been compared by
Bridges et al. (\cite{bridges91}) and Wagner et al. (\cite{wagner91}) who
performed the first CCD studies of the GCS of NGC 1399. Except for the very
inner regions, both studies found the surface density profile of the clusters
to agree well with the galaxy light.  Wagner et al. and Bridges et al. also
obtained the first color information of the GCS using the B and V bands.  
While Wagner et al. could not see a color gradient, Bridges et al. detected a
shift of 0.2 mag to the blue between 0.5\arcmin \ and 3\arcmin. The latter
authors moreover derived a turn-over magnitude of V=23.8 and a distance of 18.5
Mpc, which still is in very good agreement with modern values (Richtler et al.
\cite{richtler00}). The specific frequency (which is
the total number of globular clusters normalized to the host galaxy's
luminosity, Harris \& Van den Bergh (\cite{harris81})), derived by
Harris\,\&\,Hanes (\cite{harris87}), Bridges et al. (\cite{bridges91}) and
Wagner et al. (\cite{wagner91}) was unusually large for an elliptical galaxy
($16\pm4$), a property which is shared with other central galaxies in galaxy
clusters (e.g. Harris \cite{harris01}).

The next contribution came from the Washington system: Ostrov et al.
(\cite{ostrov93})  measured Washington colors of NGC 1399 globular clusters and
also detected the color gradient reported earlier. They also mentioned a
"multimodality" in the metallicity distribution, indicating that the GCS could
be composed of different populations with distinct properties. Kissler-Patig et
al. (\cite{kisslerpatig97})  also noted the broad V-I color distribution of the
NGC 1399 clusters in comparison with other Fornax ellipticals and a high
specific frequency, but could not confirm the color gradient seen by Ostrov et
al. (\cite{ostrov93}) and Bridges et al. (\cite{bridges91}).

An improved photometric analysis in the Washington system (Ostrov et al.
\cite{ostrov98}) showed that the color distribution of the clusters was rather
''bi-modal`` than ''multi-modal``. In this work, a rather low specific
frequency of only 5.6 was derived.

The open questions are still:  How far does the NGC 1399 GCS extend? What
substructure can be identified and how does it relate to the host galaxy? Does
the population structure change in the outer parts? How does the cluster
distribution compare to the stellar distribution? What significance does the
quoting of a general specific frequency have, if the GCS is substructured?

From the photometric side, progress can only be expected if a much larger field
is investigated than has been done in the past. With the advent of the new
MOSAIC cameras, we now have the opportunity to do this. Moreover, the excellent
metallicity resolution of the Washington photometric system enables us to
reconsider the metallicity distribution based on much better number statistics
than in previous studies.

This paper is part of a larger effort to study NGC 1399 and its cluster
population. The second paper in this series by Richtler et al.
(\cite{richtler02}) (Paper {\sc ii}) will use spectroscopic observations of
globular clusters for a dynamical study of NGC 1399.

\section{Observation \& Reduction}

\subsection{The Data and Basic Reduction}

\begin{figure}[t]
\centerline{\resizebox{\hsize}{!}{\includegraphics{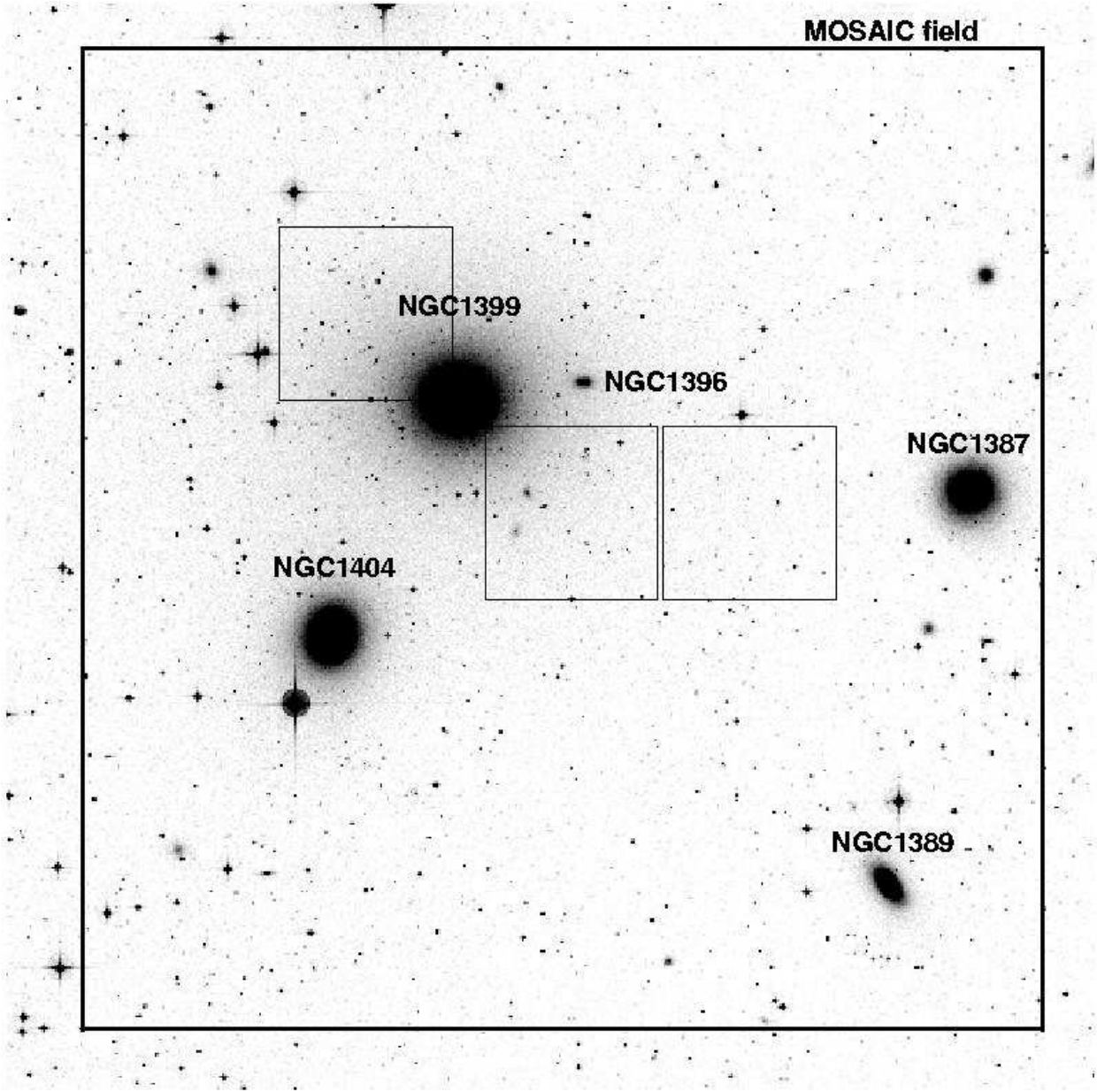}}}
\caption{
	The MOSAIC field and the three smaller VLT fields are overlayed on a DSS
        image. \label{fig:near1399fin}}
\end{figure}

Data from 5 observing runs are contributing to the present photometric study.
The main data set consists of Washington wide-field images of NGC 1399 (plus a
background field) taken with the MOSAIC camera mounted at the prime focus of
the CTIO 4m Blanco telescope. Additionally, three sub-fields were observed at
the VLT (FORS2) in V and I. The location of the MOSAIC field around NGC 1399
and the VLT fields are shown in Fig.\,\ref{fig:near1399fin} as overlays on a
DSS image.

We used the Kron-Cousins R and Washington C filters, although the genuine
Washington system uses T1 instead of R.  However, Geisler (1996) has shown that
the Kron-Cousins R filter is more efficient than T1, due to its larger
bandwidth and higher throughput, and that R and T1 magnitudes are closely
related, with only a very small color term and zero-point difference
($\mathrm{R}-\mathrm{T1}\approx -0.02$).  The MOSAIC wide-field camera images a field
of $36\arcmin\times 36\arcmin$, at a pixel scale of 0.27\arcsec/pixel. The
first three observing runs were performed in the 8 channel read-out mode; for
the last one (November 2001) the 16 channel mode was available. Further
information on the MOSAIC camera can be found in the MOSAIC homepage \footnote{
http://www.noao.edu/kpno/mosaic/mosaic.html}.

During the first run (December 1999), we observed one MOSAIC field centered
near NGC 1399 (see Fig.\,\ref{fig:near1399fin}) in the C filter, while the R band
images were kindly obtained by D.Zurek and M.Shara.

In the second run (December 2000), we observed a background field $3.5\degr$
north-east of NGC 1399 (RA$=3^h48\arcmin$ Dec$=-32\degr32\arcmin$) in C and R,
as well as Washington standard fields.

In the third run (November 2001), we observed adjacent fields to NGC\,1399 that
overlap with the central field. We also observed standard fields under
photometric conditions. This run was used for the photometric
calibration of the presented Washington data.

In addition to the MOSAIC data, we analyzed three fields ($6.8\arcmin\times
6.8\arcmin$) near NGC\,1399 for which 5\,min exposures in V and I were taken
with VLT/FORS2 in December 2000, in the course of a campaign aimed at measuring
radial velocities of globular clusters around NGC 1399 (Richtler et al.
\cite{richtler01}).

In Tab.\,\ref{tab:obssummary} we summarize the relevant information on the
different observing runs.

The MOSAIC data were reduced using the {\it mscred} package within IRAF. In
particular this software is able to correct for the variable pixel scale across
the CCD which would cause otherwise a 4\% variability of the brightness of
stellar-like objects from the center to the corners. The flatfielding resulted
in images that had remaining sensitivity variations $\le$ 1.5\%. In particular
Chip \#4 and Chip \#5 showed discernible remaining flatfield structure (but
within the given deviation).

To facilitate the search for point sources on the MOSAIC images model light
profiles of NGC\,1399 as well as of the neighboring galaxies NGC\,1404 and
NGC\,1387 were subtracted. This was done using a median filter with an inner
radius of $9.5\arcsec$ and an outer radius of $11\arcsec$. This size is large
enough for not altering the point source photometry which has been verified
with artificial star tests described later on.

\subsection{Identification of Point Sources}

We used SExtractor (Bertin \& Arnouts \cite{bertin96}) and DAOPHOT II for image
classification and photometry, respectively. Our extensive tests showed that
each program was optimized for these respective tasks. To distinguish between
point sources and extended objects, we compared the SExtractor stellarity
indices for objects found both on the MOSAIC and the VLT images. With the aid
of the superior seeing on the VLT images, we determined 0.4 as a lower limit
for the SExtractor classifier on the MOSAIC images.

Using this classification, the comparison with the VLT fields revealed that
$(90\pm5)\%$ of the point sources found in the MOSAIC fields were correctly
classified as point sources. Only $1.2\pm0.1$ galaxies $\mathrm{arcmin}^{-2}$ 
have been erroneously classified as point sources in the MOSAIC images.
With less a strict selection criterion one could lower the number of lost point
sources. However, this would come at the expense of additional contamination by
background galaxies. Since we are interested in the GCS at large radii, i.e. at
low surface densities, we rather prefer a lower completeness over an increased
contamination by galaxies.

We retained 10453 point sources while 13600 resolved objects were discarded.
The brightest non-saturated objects have R$\approx 18$ (slightly depending on
the individual MOSAIC chip).

For further study we chose a faint magnitude cut-off, depending on color. The limit
also
accounts for the fact that the background field is not as deep as the NGC 1399
field. We therefore defined this cut-off by the completeness limit of the
background field and regarded only objects brighter than this limit (see
Fig.\,\ref{fig:sel1}).

\subsection{Photometric Calibration}

Standard fields for the photometric calibration have been observed during the
November 2001 run. In each of the 3 nights, we observed 4-5 fields, each
containing about 10 standard stars from the list of Geisler
(\cite{geisler96a}), with a large coverage of airmasses (typically from 1.0 to
1.9). It was possible to use a single transformation for all three nights,
since the coefficients derived for the different nights were indistinguishable
within the errors.

We derived the following relations between instrumental and standard magnitudes: 
\begin{eqnarray} 
\nonumber
	\mathrm{T1} =
	&\mathrm{R}_\mathrm{inst}+(0.72\pm0.01)-(0.08\pm0.01)X_\mathrm{R}\\
\nonumber
		&+(0.021\pm0.004)(\mathrm{C}-\mathrm{T1})\\
\nonumber
	\mathrm{C} =	&\mathrm{C}_\mathrm{inst}-(0.06\pm0.02)-(0.30\pm0.01)X_\mathrm{C}\\
\nonumber
		&+(0.074\pm0.004)(\mathrm{C}-\mathrm{T1})
\end{eqnarray}

\begin{figure}[t]
\centerline{\resizebox{\hsize}{!}{\includegraphics{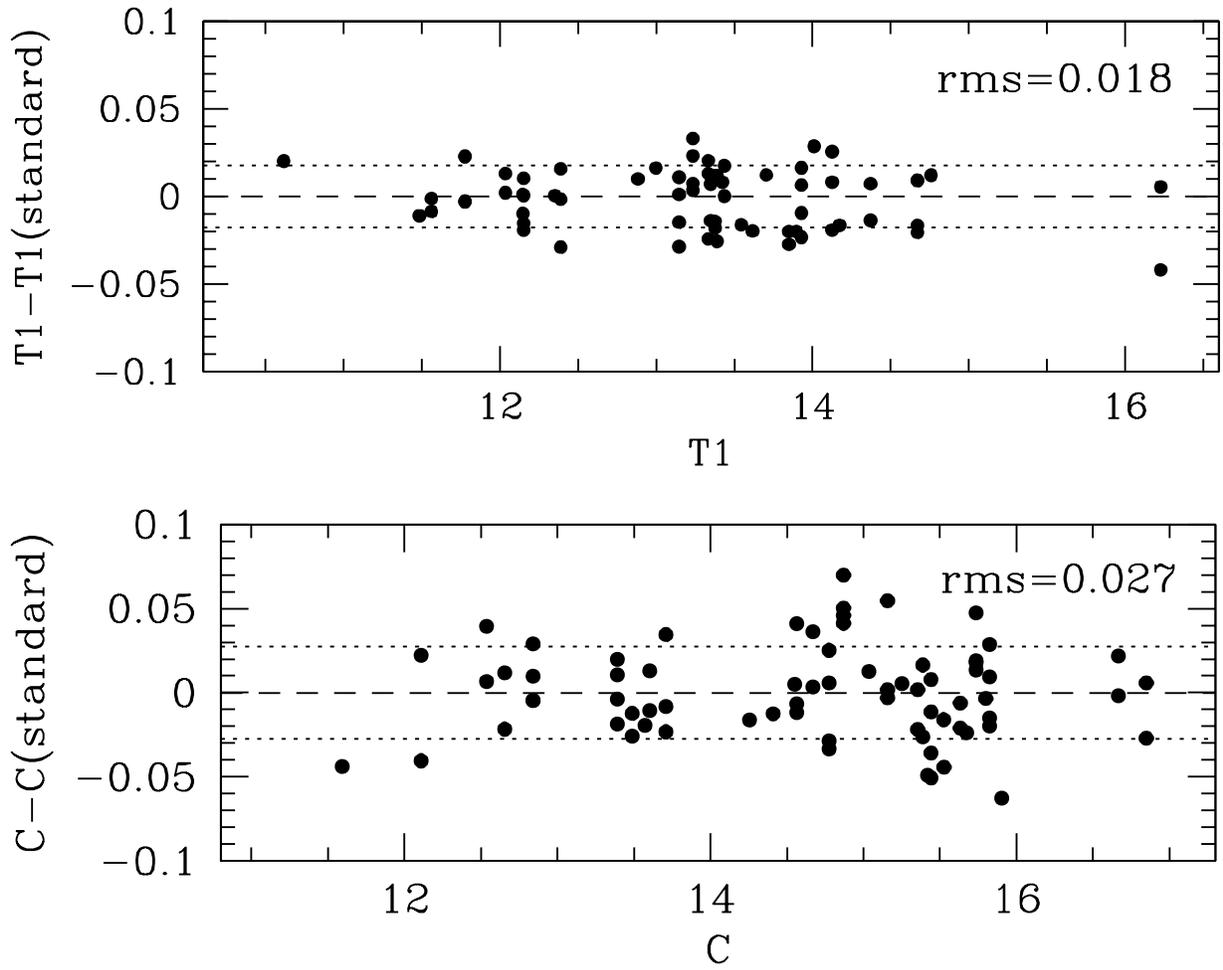}}}
\caption{
	The residuals of the standard star magnitudes after the applied calibration
	are plotted. \label{fig:calib}}
\end{figure}

The standard deviation of the difference between our calibrated and tabulated
magnitudes is 0.018\,mag in T1 and 0.027\,mag in C (see Fig.\,\ref{fig:calib}).

To calibrate the NGC\,1399 field we identified isolated stars which have been
observed on three overlapping fields during the November 2001 run. These stars
were used to determine the zero points, while the color terms were adopted from
the November 2001 run. The scatter between the zero points determined from
individual stars is 0.03\,mag and most probably due to flat field
uncertainties.

The final uncertainties of the zero points are 0.03\,mag and 0.04\,mag for R
and C, respectively. This results in a calibration uncertainty in C-T1 of
0.05\,mag (the uncertainty in the color term can be neglected).

Ostrov et al. (\cite{ostrov98}) also obtained Washington photometry for a
smaller field centered on NGC 1399. We used their Fig.\,6 and Fig.\,10 to
obtain the color and luminosity distribution of their objects. In
Fig.\,\ref{fig:colordist} we compare the color distribution of the same
subfield, and in Fig.\,\ref{fig:lumfkt_MOS} the luminosity functions are
compared. These figures show the good agreement between the two photometric
data sets.

The VLT data in V and I have been calibrated using the zero points, airmass
coefficients and color terms published on the ESO Web page for the
corresponding night \footnote{http://www.eso.org/observing/dfo/quality/
FORS/qc/zeropoints/zeropoints.html}.  We used the same zero point for all
fields despite the fact that the gain is different for all four amplifiers,
since we divided by a normalized flatfield and thus we had to average the gain
as well (in the high gain mode). The ESO calibration for the I filter includes
a color term $\mathrm{I}_0 \propto \mathrm{I}_\mathrm{inst} - 0.07
(\mathrm{V}-\mathrm{R})$.  Because we had no VLT observations in the R band, we
used a subset of clusters for which we had VLT (V,I)  and (MOSAIC) R
observations and found the the following relation between (V-I) and (V-R)
colors: (V-R)$=-(0.10\pm0.04)+(0.47\pm0.04)$(V-I), with a scatter of
$\sigma_{\mathrm{V-R}}=0.04$\,mag. This relation is used for the photometric
calibration where no R observations were available. This procedure introduces
an error of 0.04\,mag in the final magnitudes as one can judge from the scatter
of the relation. For Milky Way GCs the following can be obtained with the data
assembled in the Mc\,Master catalogue (Harris \cite{harris96}):
(V-R)$=-(0.01\pm0.01)+(0.49\pm0.01)$(V-I). Thus while the slope of the relation
is the same, the V-R color is either 0.09 mag bluer in NGC 1399 than in the
Milky Way or the V-I color is systematically 0.09 mag redder.  The reason for
this difference is unclear. We found for the three nights (the subscripts
denote the nights):

\begin{eqnarray}
\nonumber
\mathrm{V}_1 &= \mathrm{v}_\mathrm{inst}+28.057+(0.74)-0.16 X_\mathrm{V} \\
\nonumber
\mathrm{I}_1 &= \mathrm{i}_\mathrm{inst}+27.284+(0.74)-0.16 X_\mathrm{I}-0.07(\mathrm{V}_1-\mathrm{R}_1) \\
\nonumber
\mathrm{V}_2 &= \mathrm{v}_\mathrm{inst}+28.048+(0.74)-0.16 X_\mathrm{V} \\
\nonumber
\mathrm{I}_2 &= \mathrm{i}_\mathrm{inst}+27.263+(0.74)-0.16 X_\mathrm{I}-0.07(\mathrm{V}_2-\mathrm{R}_2) \\
\nonumber
\mathrm{V}_3 &= \mathrm{v}_\mathrm{inst}+28.038+(0.74)-0.16 X_\mathrm{V} \\
\nonumber
\mathrm{I}_3 &= \mathrm{i}_\mathrm{inst}+27.261+(0.74)-0.16 X_\mathrm{I}-0.07(\mathrm{V}_3-\mathrm{R}_3) 
\end{eqnarray}

We note the term due to the conversion from ADUs to electrons (0.74) separately
to facilitate the comparison with the zero points given in the ESO Web page.

For the reddening towards NGC\,1399 we adopted, according to Schlegel et al.
(\cite{schlegel98}), E$_{\mathrm{B-V}}=0.013$. Using
E$_{\mathrm{C-T1}}=1.97$\,E$_{\mathrm{B-V}}$ (Harris\,\&\,Canterna
\cite{harris77}) we have to correct C-T1 by 0.026\,mag. For the comparison
field the foreground reddening is negligible: E$_{\mathrm{B-V}}=0.002$.

The color magnitude diagrams for the point sources in the MOSAIC images are
plotted in Fig.\,\ref{fig:sel1}. The combined color-magnitude diagram for the
three VLT fields is shown in Fig.\,\ref{fig:cmdVLT}.

\subsection{Photometric Completeness}

The globular cluster luminosity function (GCLF) will be derived from the VLT
images only, so the photometric incompleteness as a function of apparent
magnitude of the MOSAIC images is not of much interest to our current study.  
However, it is interesting to know the global incompleteness, and how many
objects, being resolved in the VLT images, were falsely classified as point
sources in the MOSAIC images.  We found that within our magnitude range,
$(41\pm3)\%$ of the cluster candidates visible on the VLT frames were detected
on the MOSAIC images. 0.5\,objects/arcmin$^2$ which are resolved in the VLT
images have been classified as point sources in the MOSAIC images
(objects brighter than the employed faint magnitude limit).
This is a reassuringly small number.

Dithering to fill the gaps between the eight individual CCD chips was done only
for the C images.  Therefore one has to consider the spatial incompleteness due
to the gaps between the MOSACI CCDs in the R images, saturated galaxy centers
and some bright stars west of NGC 1399 which account for a spatial
incompleteness of $\approx 4\%$ in the MOSAIC data.

Experiments with artificial stars were performed in order to check on whether a
magnitude-dependent difference between input and output magnitude is observed
among the point sources. This was not the case, and thus the size of the median
filter we used to model and subtract the galaxy light, is justified.  
Additionally, we found that for point sources more distant from NGC\,1399 than
$1.6\arcmin$, no radially varying completeness has to be corrected for.

The completeness of the VLT fields has been determined with extensive
artificial star tests. We used {\it addstar} of the DAOPhot package under IRAF
to add 5 times 2000 stars to each VLT image. The stellar PSF was the same as
determined by the PSF fitting routine in the same package. The color and
magnitude range encompassed the range of observed globular clusters. The
resulting completeness as a function of V magnitude is shown in
Fig.\,\ref{fig:VLTcompl}.

\section{The Color Magnitude Diagram}

\begin{figure}[t]
\centerline{\resizebox{\hsize}{!}{\includegraphics{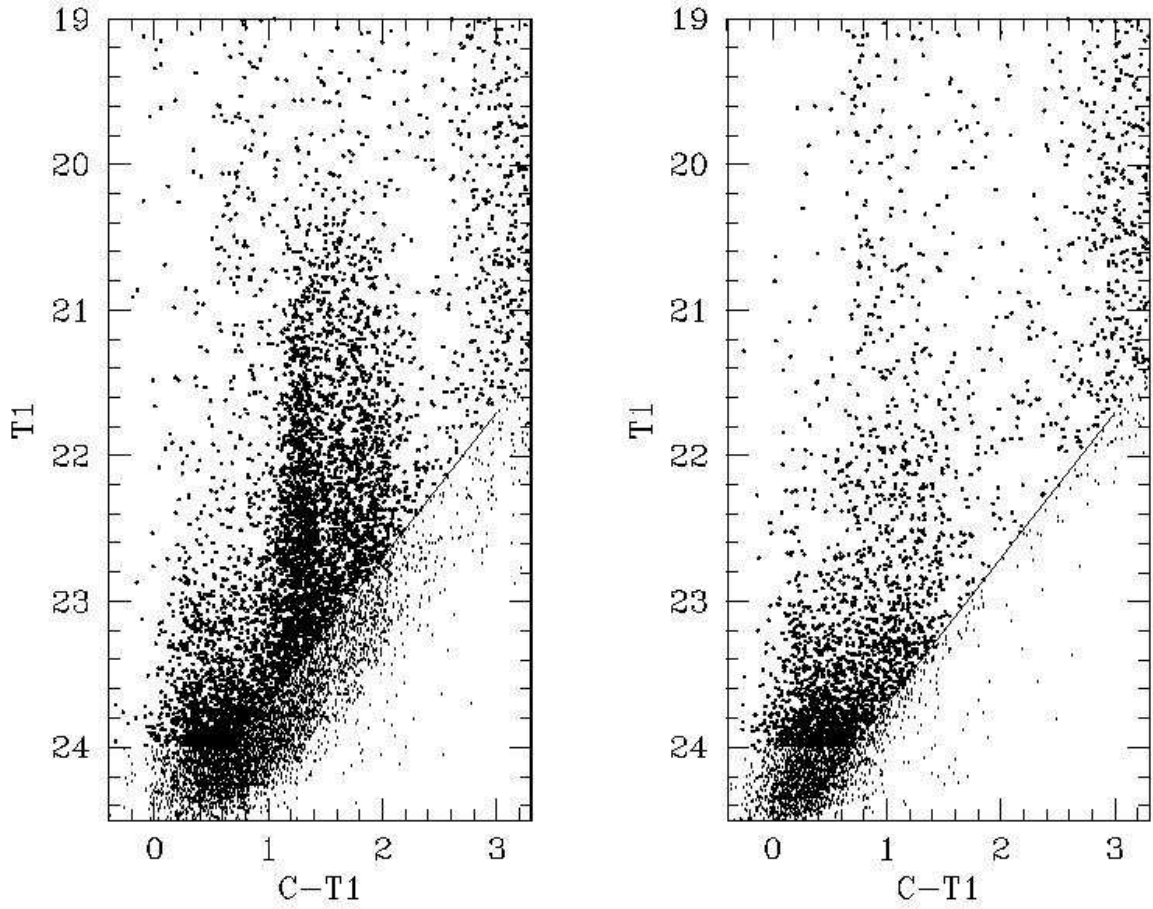}}}
\caption{
	The CMD based on the MOSAIC data is shown in the left panel for
        the NGC\,1399 field, and in the right panel for the background field.
        The line indicates the limiting magnitude which is used throughout the paper.
        Brighter stars are also plotted with slightly thicker dots than stars fainter
	than the limiting magnitude. \label{fig:sel1}}
\end{figure}

\begin{figure}[t]
\centerline{\resizebox{\hsize}{!}{\includegraphics{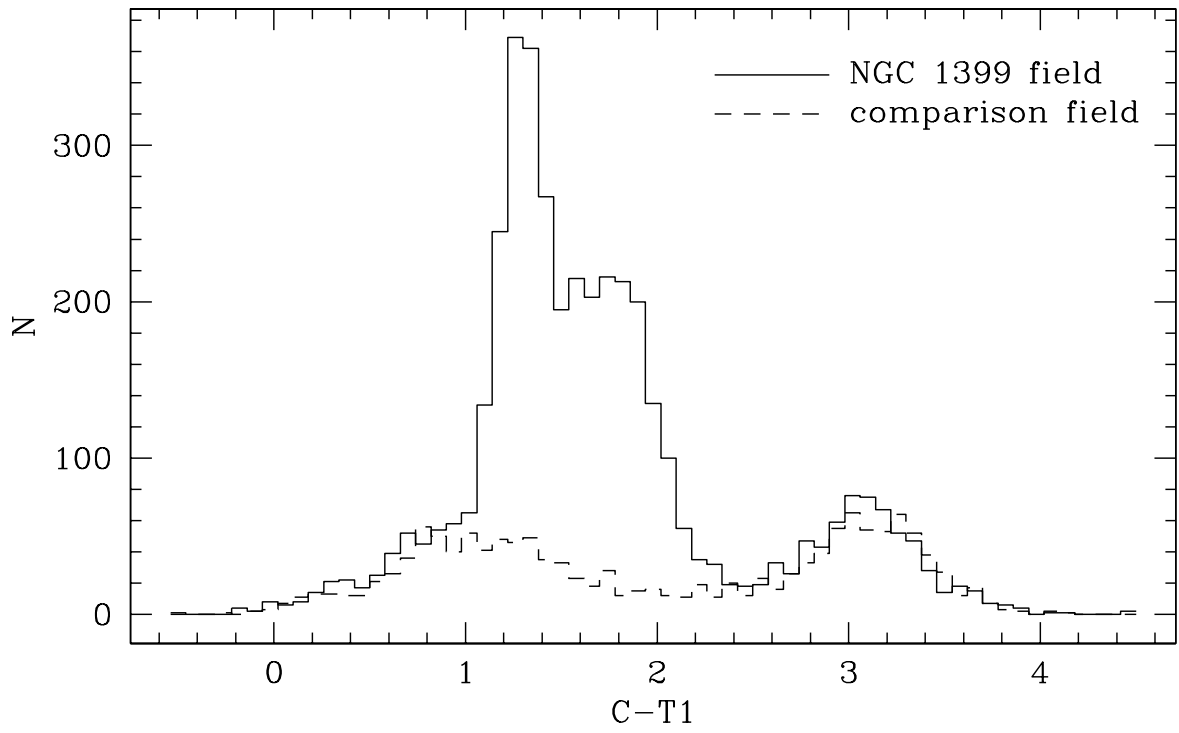}}}
\caption{
	Comparison of the color histograms of all selected point sources of the NGC 1399 field
        (solid line) and the comparison field (dashed line). \label{fig:colorhistall}}
\end{figure}

\begin{figure}[t]
\centerline{\resizebox{\hsize}{!}{\includegraphics{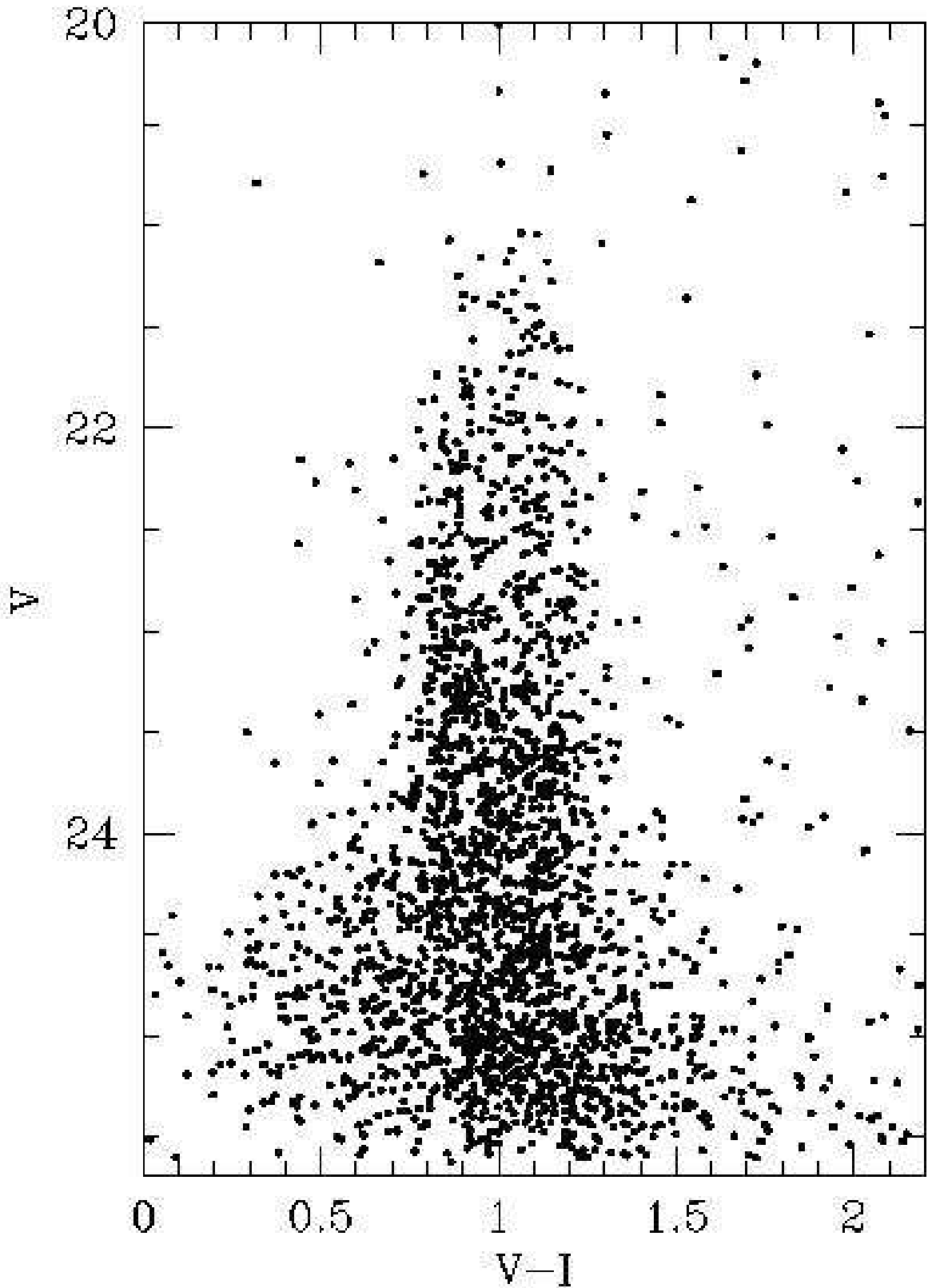}}}
\caption{
	The CMD assembled from the three VLT fields.The bimodality of the
        color distribution is visible, although not as nicely as in the Washington
        colors (Fig.\ref{fig:sel1}). We shall use the VLT photometry mainly
        for deriving the globular cluster luminosity function. \label{fig:cmdVLT}}
\end{figure}

\begin{figure}[t]
\centerline{\resizebox{\hsize}{!}{\includegraphics{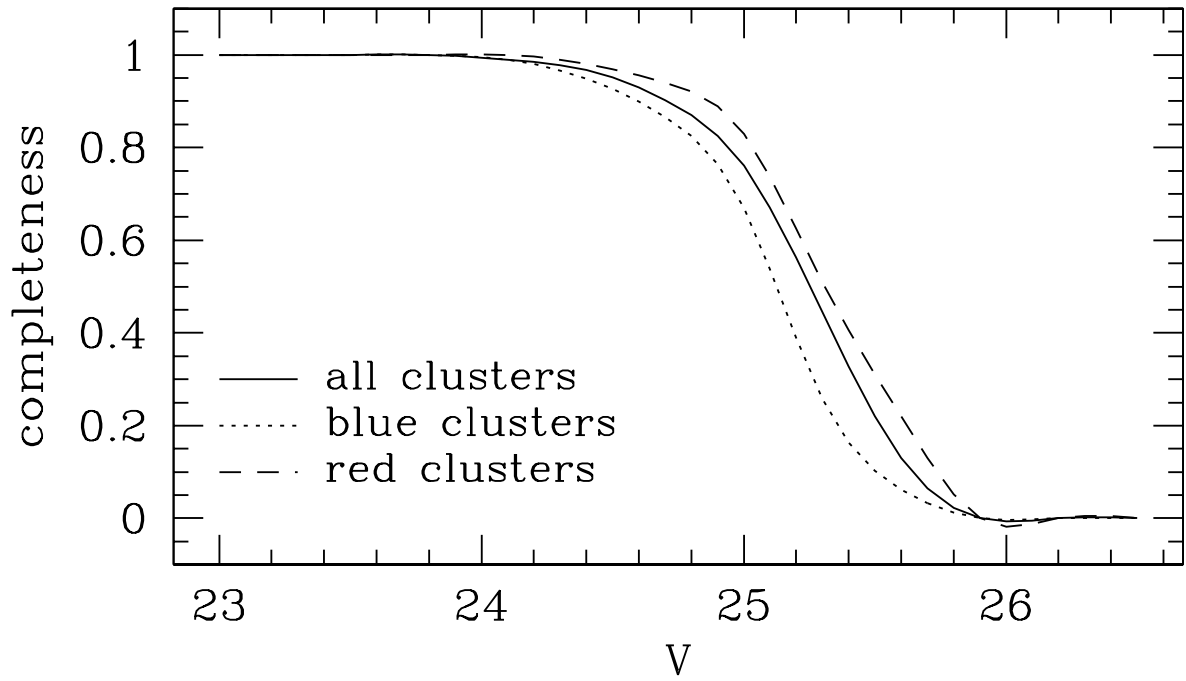}}}
\caption{
	Completeness function for the VLT data (Field 1). \label{fig:VLTcompl}}
\end{figure}

We will now discuss the CMD presented in Fig.\,\ref{fig:sel1} and
Fig.\,\ref{fig:cmdVLT}. The globular clusters show up nicely in
Fig.\,\ref{fig:sel1} within the color range $1.1<\mathrm{C-T1}<2.3$. In its
vast majority the cluster population belongs to NGC 1399, but the GCSs of NGC
1404 and NGC 1387 are also included. The combination of a large area, depth,
and broad color baseline render this CMD unique among the available photometric
data sets of cluster systems. The cluster population starts at about T1=20.
Well visible is the blue ''peak'' in the color distribution, while the red
clusters rather build a ''shoulder'' and do not distinctly peak at a certain
color (see section 5 for a detailed discussion of the color distribution). The
comparison with the background field (Fig. \ref{fig:colorhistall}) shows how
well it reproduces the shoulder starting at C-T1 = 0 and the red peak (stars) at
C-T1 = 3. 

\begin{figure}[t]
\centerline{\resizebox{\hsize}{!}{\includegraphics{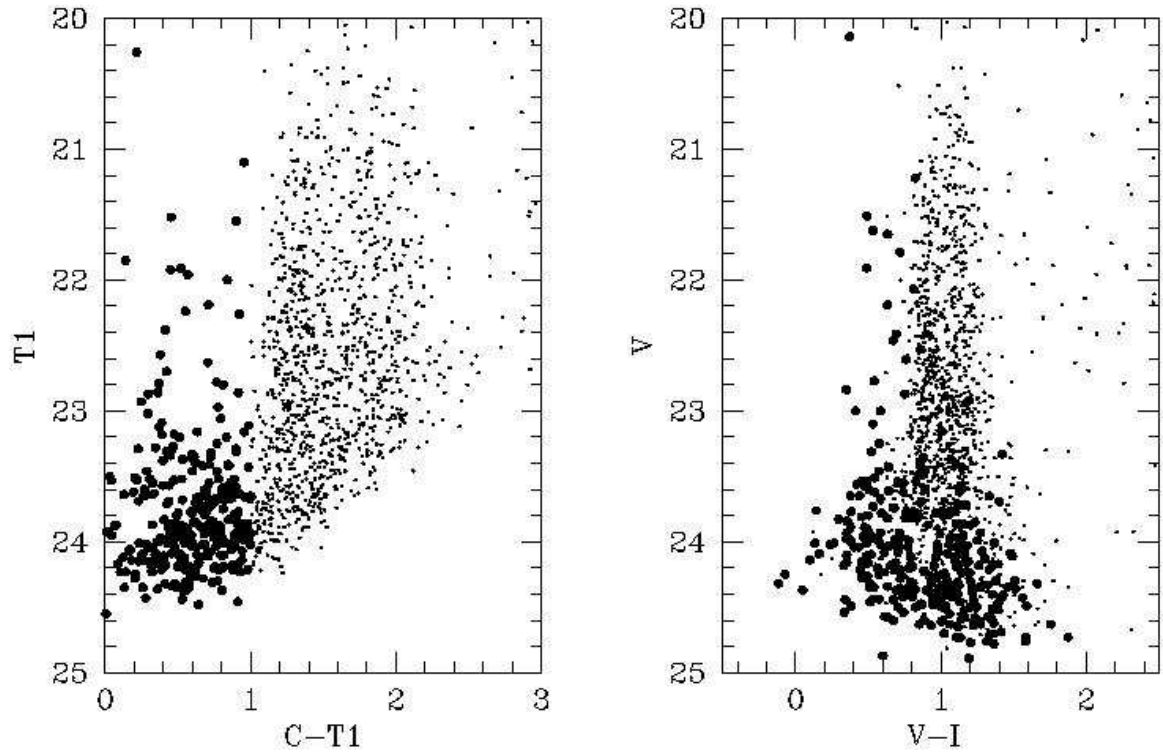}}}
\caption{
	The CMDs of point sources that have been
        identified on both the VLT and the MOSAIC images. We marked objects
        bluer than C-T1=1 by filled circles.These are predominantly
        galaxies. Note how these objects contaminate the CMD in V,V-I. \label{fig:wash_sel}}
\end{figure}

The comparison between the VLT and the MOSAIC data also shows why the
contamination of the cluster system by background galaxies is much lower for
the filter combination C,R than for V,I, even if the seeing is considerably
worse. The reason can be seen in Fig.\,\ref{fig:wash_sel}. The main sources
that contaminate the cluster sample in V \& I are unresolved faint blue galaxies that have
an UV excess and are thus bluer than cluster candidates in C-R while having the
same V-I color. We confirm the galaxy nature of these (C-T1)-blue objects since
they have a homogeneous distribution and are not concentrated towards NGC\,1399
like the cluster candidates.

\section{The Globular Cluster Luminosity Function}

To derive the globular cluster luminosity function (GCLF) of the VLT data we selected
clusters within a radial distance of 4.3\arcmin \ from the center of NGC\,1399.  
As a background field we used the FORS2 image centered 13\arcmin \ away from
NGC\,1399. We are aware of the fact that this ''background field'' still
contains a considerable number of clusters apart from background objects, which
deteriorates the statistics, but should not alter the location of the turn-over
magnitude (TOM). 

\begin{figure}[t]
\centerline{\resizebox{\hsize}{!}{\includegraphics{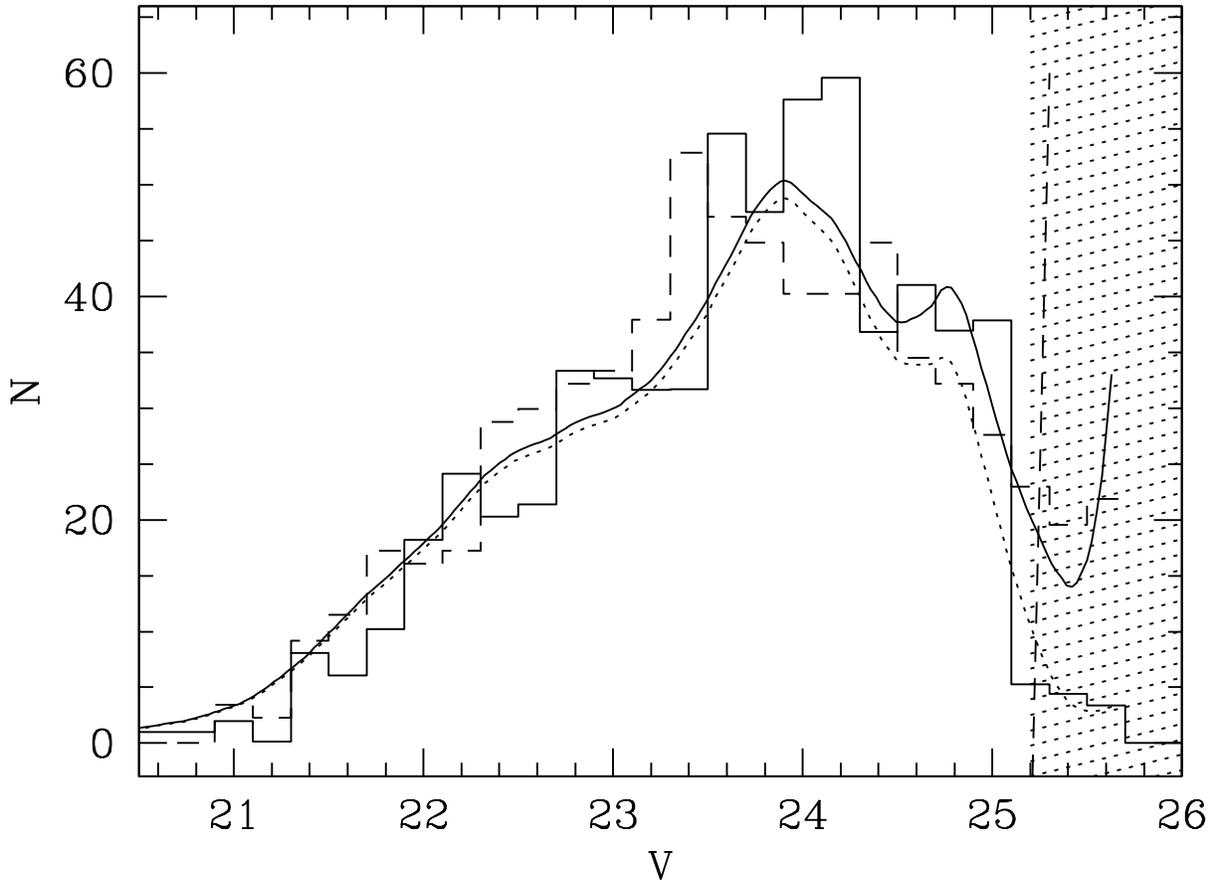}}}
\caption{
	The luminosity function of the cluster candidates within a
        adial range of 4.3\arcmin \ in the FORS2 images 
        is plotted as solid line. The shaded area designates the
        magnitude range beyond the 50\% completeness. The dotted line shows the
        histogram before the completeness correction. The dashed histogram 
        indicates the (arbitrarily scaled) HST GC luminosity function of
	Grillmair et el. (\cite{grillmair99}). \label{fig:lumfkt_VLT}}
\end{figure}

\begin{figure}[t]
\centerline{\resizebox{\hsize}{!}{\includegraphics{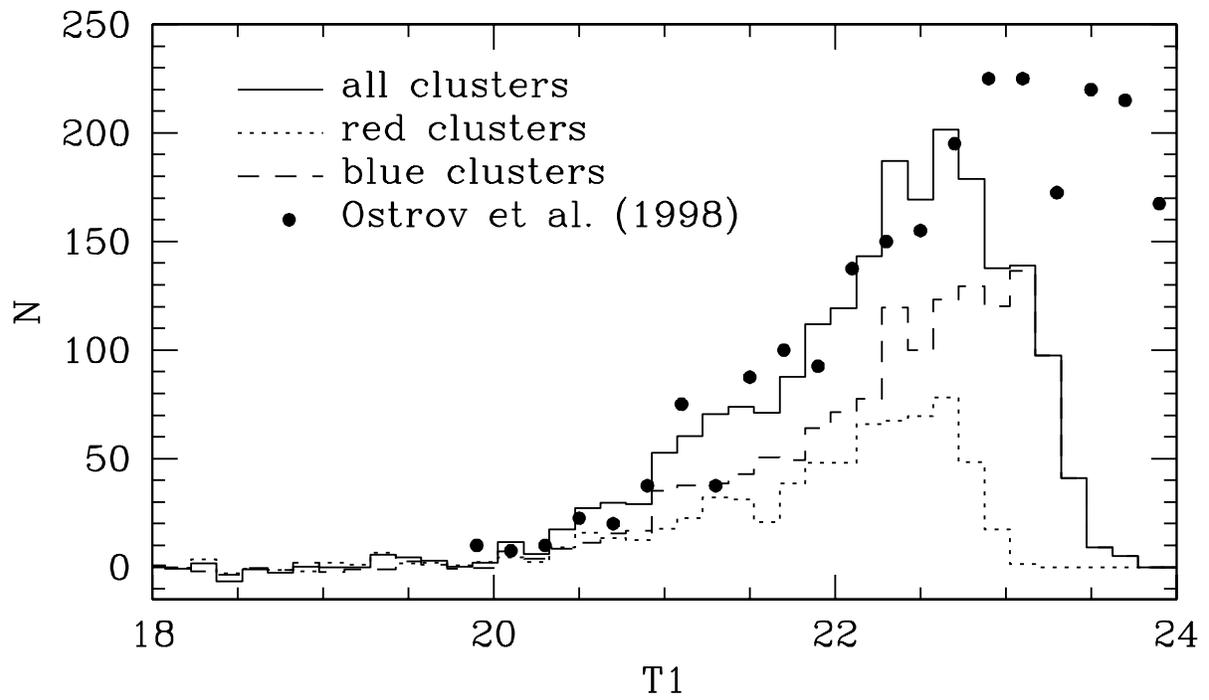}}}
\caption{
	The luminosity function of the cluster candidates from the
        MOSAIC data is shown for the whole sample (solid line), and the
        blue and red subsamples (dotted and dashed line, respectively).
        The luminosity function given by Ostrov et al. (\cite{ostrov98})
        is also shown (arbitrarily scaled). \label{fig:lumfkt_MOS}}
\end{figure}

The luminosity function is also plotted as kernel filtered density
distributions. We followed here the approach described by Merritt\,\&\,Tremblay
(\cite{merritt94}) and used an adaptive Epanechikov kernel. The scaling in
Fig.\,\ref{fig:lumfkt_VLT} is number of clusters per 0.2 magnitude. In the
following, all distributions using this nonparametric estimation are based on
the Epanechikov kernel. For the fitting we use both, filtered data and the
histograms to ensure that the results do not depend on the representation.

We fitted a Gaussian to the distributions for different limiting magnitudes
(always brighter than the 50\% completeness) and found an average
$\sigma=1.30\pm0.12$ and $\mathrm{m}_\mathrm{TO}=24.0\pm0.1$. 
However, a correlation between
$\sigma$ and TOM exist these fits which is caused by the fact that our
observations do not extend far enough beyond the TOM. It is therefore advisable
to keep the width fixed and to fit only the TOM. With a $\sigma=1.2, 1.3,
1.4$, which are used for elliptical galaxies (Ashman \& Zepf \cite{ashman98}), we
find a TOM of $23.98\pm0.11, 24.02\pm0.09, 24.05\pm0.12$. We
choose the same final TOM of $\mathrm{m}_\mathrm{TO}=24.0\pm0.1$.

With the knowledge of the TOM we calculate that the fraction of all clusters
which are found in the area covered by the VLT data is $85\%\pm4\%$.

The background-subtracted luminosity functions of the MOSAIC cluster candidates
from 2\arcmin \ to 16\arcmin \ (excluding 2\arcmin \ around NGC 1404) are shown
in Fig.\ref{fig:lumfkt_MOS}. The luminosity distributions are consistent with
the assumed distance modulus of $\mathrm{m}-\mathrm{M}=31.4$ and a width of the
Gaussian luminosity function of 1.2.

This TOM together with the known completeness of the VLT data can be used to
determine the completeness of the MOSAIC fields. Since $41\%\pm3\%$ of the VLT
candidates were found on the MOSAIC images (see above) the absolute
completeness is $36\%\pm5\%$ for the MOSAIC images.

An alternative approach to find the completeness is to use the luminosity
function of the MOSAIC clusters directly: using a TOM of T1$=23.3$ and a width
of $\sigma=1.2$ (Forte et al. \cite{forte01}) we find a completeness of $41\pm
2\%$. We adopt a final completeness of $34\%\pm5\%$, taking into account a
spatial incompleteness due to the undithered gaps of $4\%$.

The GCLF is frequently used as a distance indicator for early-type galaxies due
to its universal peak luminosity (e.g. Harris \cite{harris91}, Harris \cite{harris01}, 
Ashman et al. \cite{ashman95}, Kundu \& Whitmore \cite{kundu01a}, Larsen et al.
\cite{larsen01}). However, due to the uncertainty of the VLT calibration we do
not use the obtained TOM for more than the above presented completeness
calculations for which the absolute calibration is not critical.

In the following we use a distance to NGC\,1399 of
$\mathrm{m}-\mathrm{M}=31.4\pm0.2$ consistent with SBF distances
(Tonry et al. \cite{tonry01}, Liu et al. \cite{liu02}),
however only marginally consistent with the result of Grillmair et al.
(\cite{grillmair99}), from the GCLF technique.

\clearpage

\section{Color Distribution of the Clusters}

In the CMD of all point sources around NGC 1399 (Fig.\,\ref{fig:sel1}) the GC
candidates stand out strikingly. A blue peak at C-T1 = 1.3 followed by a broad
distribution towards redder colors is prominently visible.  In the following,
we define GC candidates by the magnitude interval $20 < \mathrm{T1}< 23$ and
the color interval $0.8 < C-T1 < 2.3$.

\begin{figure}[t]
\centerline{\resizebox{\hsize}{!}{\includegraphics{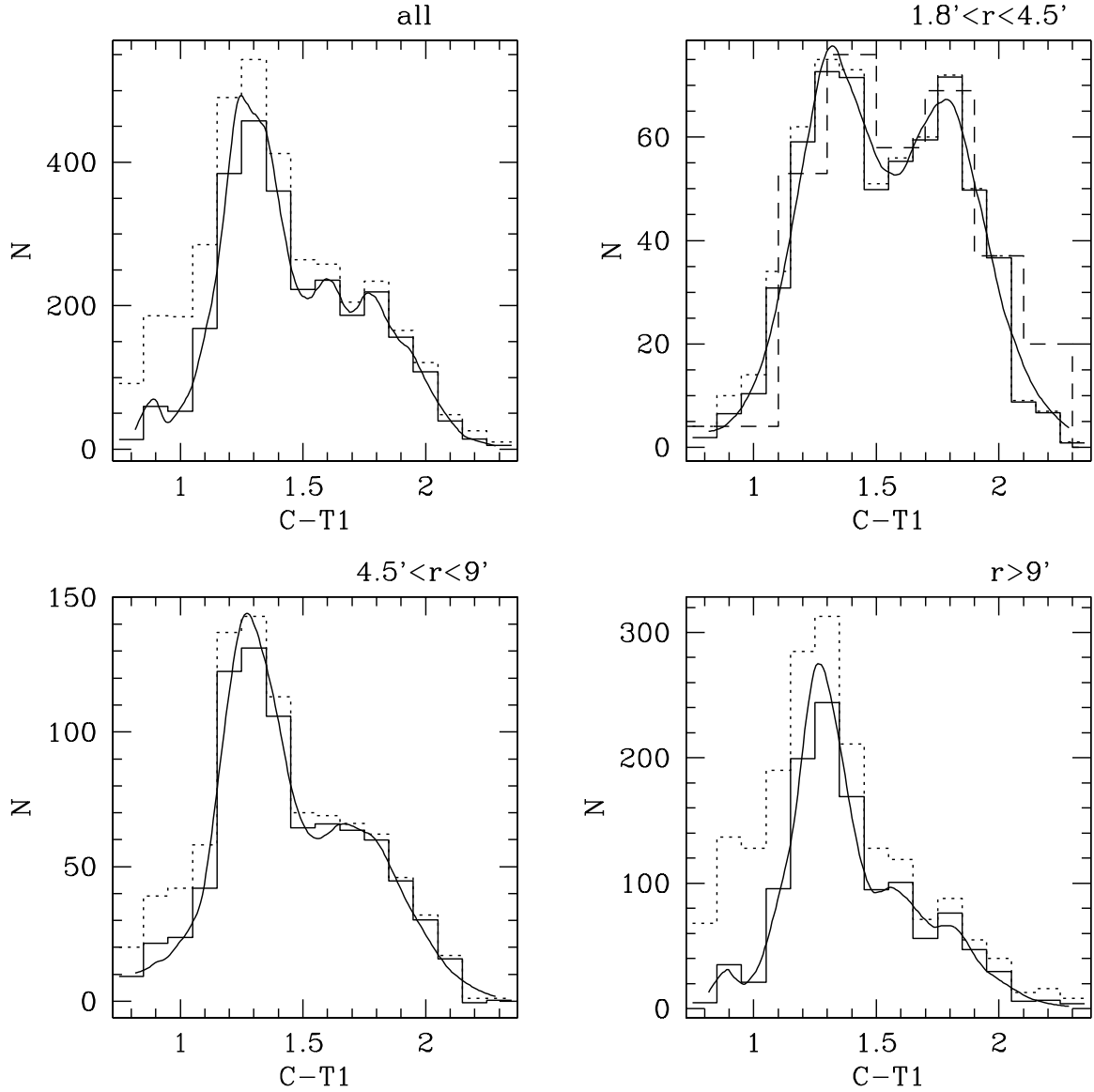}}}
\caption{
	The color distribution of globular clusters is plotted for
        the entire sample and, for three samples within
        different radial intervals which are indicated at the upper right of
        each panel. The dotted histogram shows the distribution
        before the statistical background subtraction. The smooth line is
        the adaptive kernel filtered distribution (described in Sect.4). The dashed histogram in the
        upper right panel shows the color distribution by Ostrov et al.
        (\cite{ostrov98}) within the same area and demonstrates the good agreement of the
        photometry. The counts of Ostrov et al. have not been scaled. For an interpretation of
        these observed distributions see Sect.\,5 of the text. \label{fig:colordist}}
\end{figure}

In Fig.\,\ref{fig:colordist} we plot the background subtracted color
distribution for the cluster candidates for the entire sample and in three
different ranges in radial distance. It is apparent that the color
distribution is radially dependent: at larger radii the relative contribution
of the red clusters is much smaller than at lower radii. This already shows
that the blue clusters exhibit a shallower density profile than the red
clusters, as has already been found by several authors. However, it is also
apparent that this might not hold for larger radii: the two outer distributions
(the two lower panels in Fig.\,\ref{fig:colordist}) look quite similar and are
statistically not distinguishable. The distributions plotted in
Fig.\,\ref{fig:colordist} are tabulated in Table\,\ref{tab:colordistr}. The two
peaks in the innermost sample are at $\mathrm{C}-\mathrm{T1}=1.32\pm0.05$ and
$1.79\pm0.03$, respectively.

Since so far only a few GCS have been studied in the Washington system, our
possibilities for comparison with other work are quite limited.

As mentioned above, NGC\,1399 has already been studied by Ostrov et al.
(\cite{ostrov98}) who found a bimodal distribution with peaks at C-T1$\approx
1.4$ and C-T1$\approx 1.8$, agreeing with our values for the inner region. This
is also shown in the upper right panel of Fig.\,\ref{fig:colordist}.

\section{Radial Distribution of the Clusters}

The area we cover is the largest ever observed for the NGC 1399 GCS. However,
the innermost region is unaccessible due to the brightness of the galaxy and
the saturated parts of the images. Therefore we combined our data with HST
observations by Forbes et al. (\cite{forbes98}) in order to include also the
inner part of NGC 1399 in our study. We extrapolated all three samples (HST,
VLT, Mosaic) to the entire luminosity function with the parameters given in
section 3. In Fig.\,\ref{fig:rad_distr_MOS_VLT_HST} the resulting surface
densities of GCs are plotted for the three samples.  The HST results are in
excellent agreement with our data at overlapping radii, demonstrating the
validity of our completeness calculations. A fit which satisfactorily
reproduces the surface densities of the total number of clusters after
completeness correction is:

\begin{equation}\nonumber
n(r) = 200\cdot\left(1+(r/1\arcmin)^2\right)^{-0.8}[\mathrm{arcmin}^{-2}]
\end{equation}

In the following, we focus on the MOSAIC sample. From the color histograms in
Fig.\,\ref{fig:colordist} one can immediately see that the red clusters are
more centrally concentrated than the blue ones. Guided by
Fig.\,\ref{fig:colordist}, we define the color limit between red and blue
clusters to be C-T1 = 1.55, which is the minimum seen in the 
clearly bimodal color distribution seen in the upper right panel of
Fig.\,\ref{fig:colordist}. We can now quantify the radial surface density
distributions displayed in Fig.\ref{fig:GCslope}, which show the density
profiles for both populations. Because the inner region, where a core is
visible, is excluded, we can describe the density profiles by pure power-laws.

Out to a galactocentric radius of about $7\arcmin$, the blue clusters show a
distinctly shallower slope than the red clusters. This has already been found
in earlier studies of NGC 1399 (e.g. Forbes et al. \cite{forbes98}, Ostrov et
al. \cite{ostrov98}) as well as in other galaxies, NGC 4636 being the most
striking example (Kissler et al. \cite{kissler94}).

The region beyond $7\arcmin$ has as yet been unexplored and indeed we see an
interesting change. The distribution of the red population is well represented
by a single power-law over the full radius interval between $2\arcmin$ and
$20\arcmin$, but the blue population shows a change from a shallow profile in
the inner region to a steeper profile in the outer region. Here, two power-laws
are more appropriate. For the blue clusters, the profiles are proportional to
$r^{-0.8\pm0.17}$ and $r^{-1.71\pm0.21}$ (for $1.5' < r < 7'$ and $7'<r<21'$,
respectively). For the red population, we find the profile $\propto
r^{-1.64\pm0.10}$. Thus we cannot see any significant difference in the slope
between red and blue cluster distributions beyond $7'$ of radial distance.
This can also be quantified with a K-S test for which we selected
only clusters brighter than R=22.5 to minimize the effect of the background.
This test results in a 48\% probability that
the red and blue clusters at large radii have a different radial distribution.
This probability should be considered as upper limit, since we did not correct 
for the background contamination which affects
the red and blue color range differently. This indicates again that the distribution 
of red and blue clusters is consistent with being the same.

One might recognize that the slopes fitted to the MOSAIC data at large radii
are shallower than the fit to all GCs over the whole radial range presented
above. This is also apparent from Fig.\,\ref{fig:rad_distr_MOS_VLT_HST}, which
suggests that the assumption of a uniform power-law over the whole
galactocentric distance range might not be adequate. At larger radii -- where
the investigation is restricted to the MOSAIC data -- the slope is shallower.

To provide an independent check, we derived the radial profiles for the red and
blue subsamples between $1'$ and $8'$ also from the VLT data. We only used
cluster candidates brighter than V=22\,mag to avoid the background
contamination coming from the faint blue galaxies, which in V-I cannot be
distinguished from cluster candidates (see Fig.\,\ref{fig:wash_sel}). Moreover,
the VLT fields only cover approximately $11\%$ of the MOSAIC field and thus the
statistics are poorer and the errors larger. In spite of this, the resulting
density distributions of metal-poor ($0.7<(V-I)<0.95$) and metal-rich clusters
($1.05<(V-I)<1.2$) agree very well with the results obtained with the
Washington observations: The exponents are $-1.0\pm0.13$ and $-1.5\pm0.18$ for
the blue and the red populations, respectively. However, a difference emerges
for the blue population at small radii. If we only consider the blue clusters
within $5'$, we obtain an exponent of $-0.5\pm0.09$ compared to $-1.2\pm0.2$
(based on 4 points) outside this radius. This indicates that the assumption of
two power-laws describing the metal-poor population is probably too simple.

Besides NGC\,1399, NGC\,4472 -- the brightest galaxy in the Virgo cluster -- is
the only elliptical galaxy where the GCS has been studied with wide-field CCD
photometry (Rhode\,\&\,Zepf \cite{rhode01}). Due to its similar distance (17
Mpc) and similar size the results for this galaxy can readily compared with
ours. In particular the radial distributions of red and blue clusters found by
Rhode\,\&\,Zepf (\cite{rhode01}) is very similar to our results for NGC\,1399.
This includes the mentioned ''out-leveling'' of the number ratio of red versus
blue clusters at large radii, i.e. the same radial decline of red and blue
clusters is indicated.

\subsection{The Brightest Clusters}

\begin{figure}[t]
\centerline{\resizebox{\hsize}{!}{\includegraphics{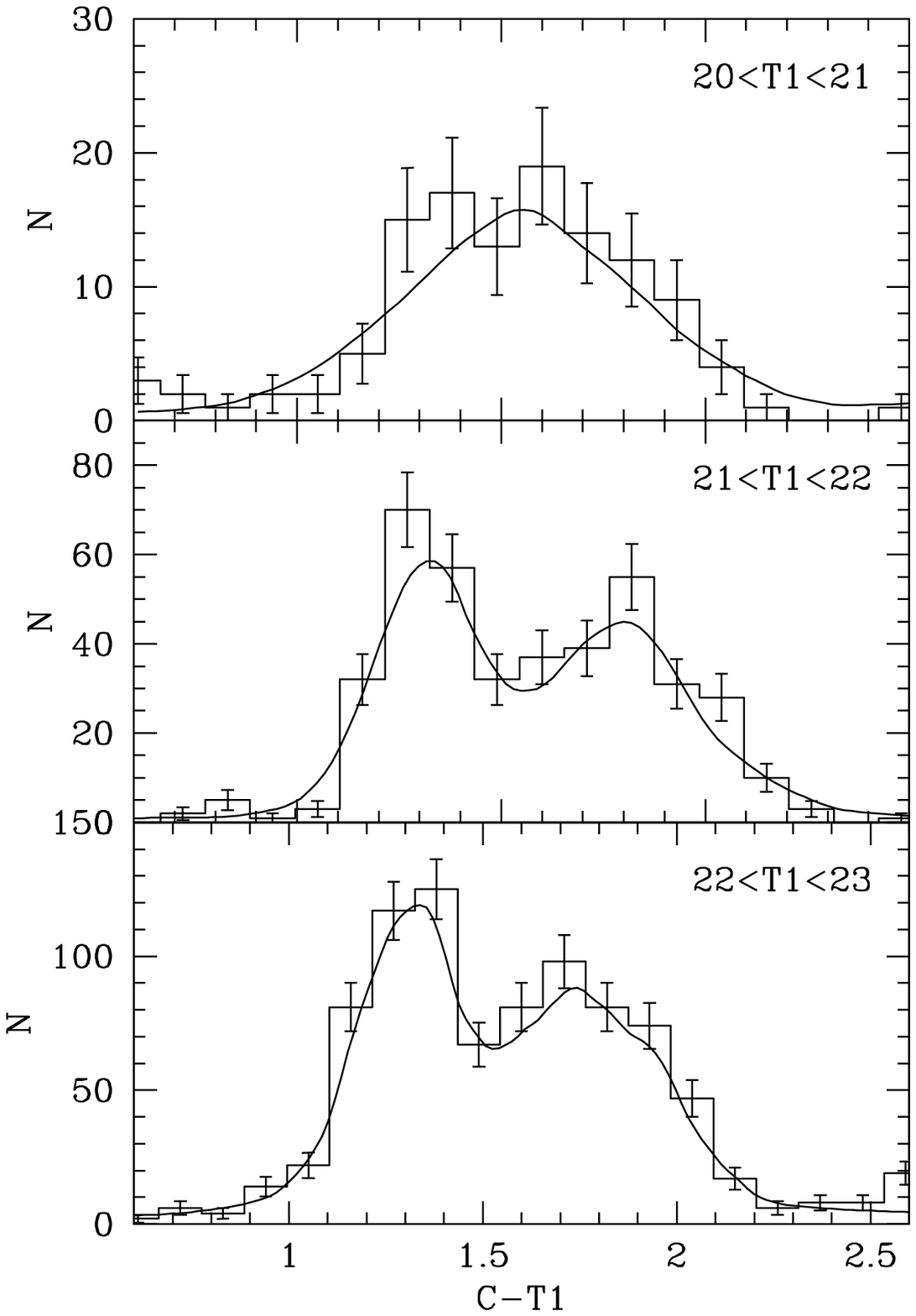}}}
\caption{
	The color distribution of globular clusters is plotted for
        three different luminosity intervals. The brightest clusters do
        not exhibit a bimodal pattern. \label{fig:color_distr_colordep}}
\end{figure}

\begin{figure}[t]
\centerline{\resizebox{\hsize}{!}{\includegraphics{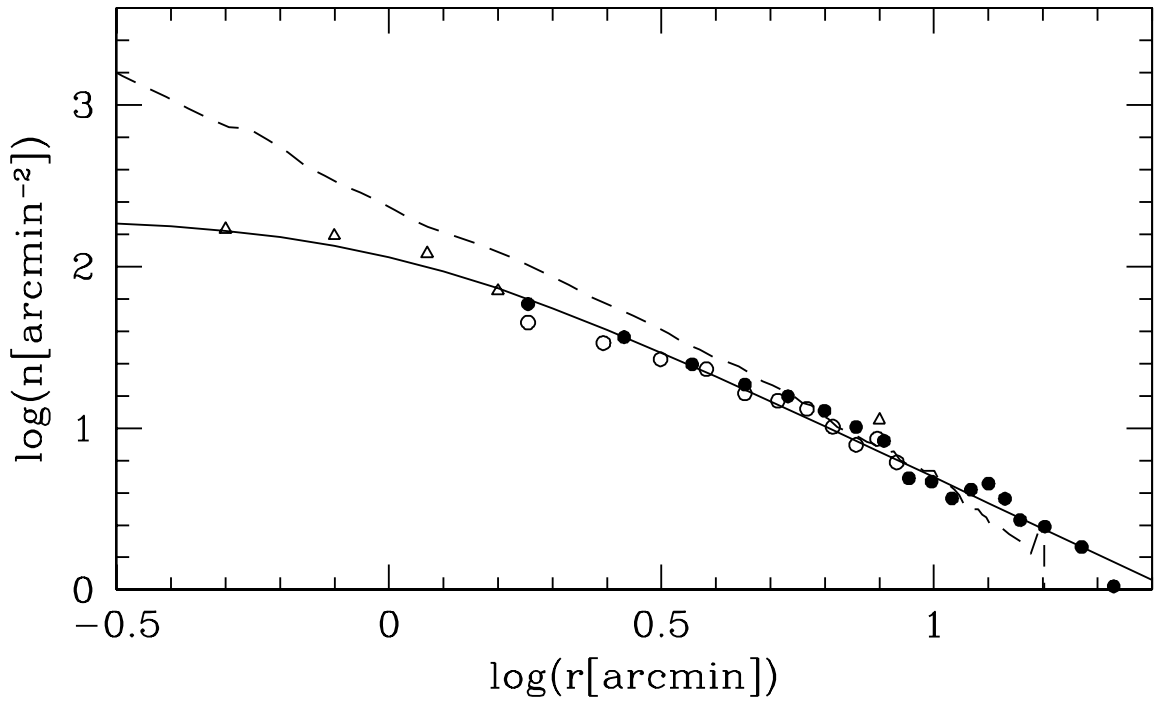}}}
\caption{
	This plot shows the radial density distribution of all clusters
        (extrapolated to the entire luminosity function) derived from
        the MOSAIC data (open circles), the VLT data (filled circles) and
        from HST data (triangles, Forbes et al. \cite{forbes98}). The solid line is
        the fit to the data. The dashed line shows the luminosity profile of NGC 1399,
        which is described in Sec.7. \label{fig:rad_distr_MOS_VLT_HST}}
\end{figure}

\begin{figure}[t]
\centerline{\resizebox{\hsize}{!}{\includegraphics{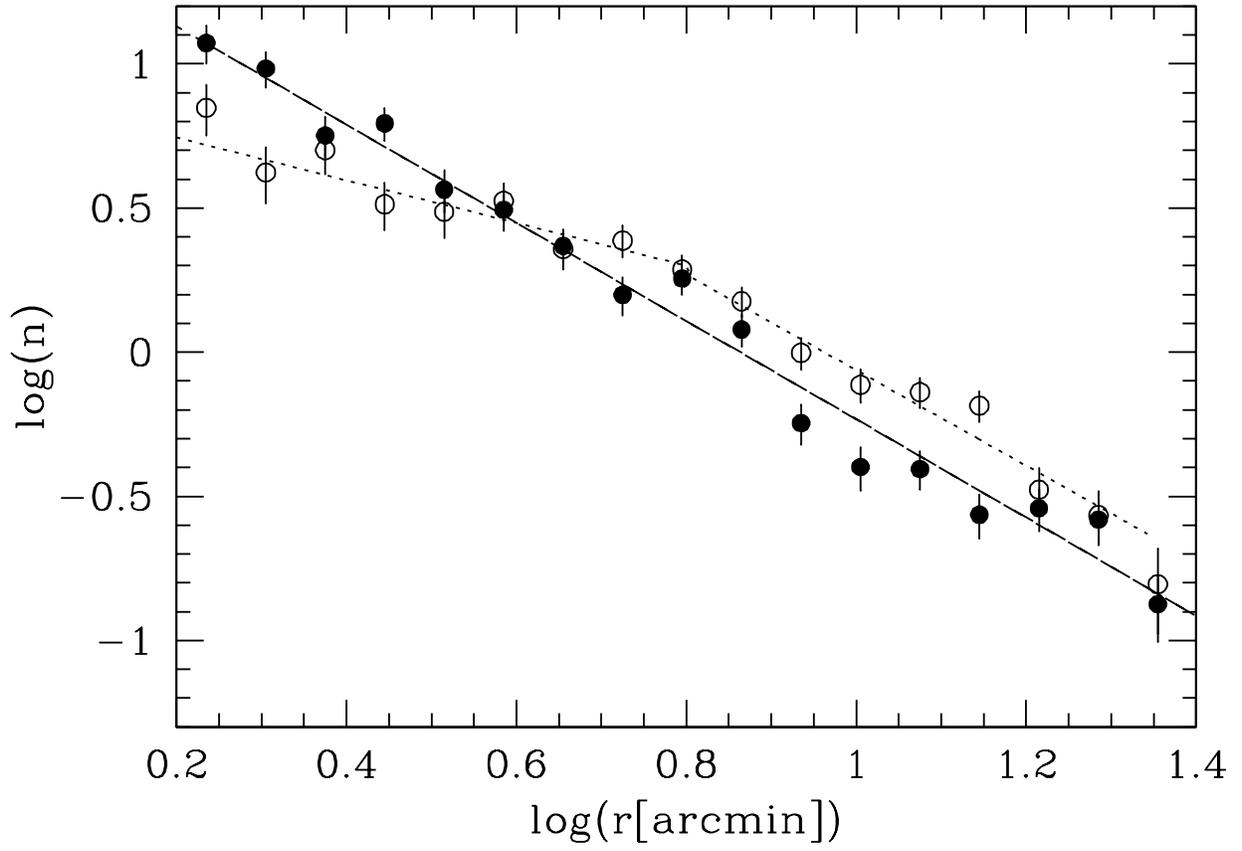}}}
\caption{
	The radial density profiles of the observed number of red (filled circles)
        and blue (open circles) globular clusters detected on the MOSAIC image are
        plotted together with the fitted power-laws. The counts are tabulated in
        Tab.\ref{tab:raddistr}. \label{fig:GCslope}}
\end{figure}

In Fig.\,\ref{fig:color_distr_colordep} we show the color distribution for
three different brightness bins. It is striking that for the brightest
clusters, shown in the uppermost panel, the color distribution differs 
largely from the fainter samples
seen (this is already apparent in the CMD): compared to the fainter sample 
mor intermediate colored ($\mathrm{C-T1}\approx1.55$) are present.
A Kolmogorov-Smirnov test 
(K-S test) resulted in only a 5\% probability that the bright sample has 
the same color distribution as the fainter one (R=21-22).
Ostrov et al. (\cite{ostrov98})
suggested that the brightest clusters might have a unimodal distribution,
however their lower number of objects did not permit a clear statement. 
In the
faintest magnitude bin, the red peak is shifted towards the blue with respect
to the intermediate bin. This, however, may be easily explained by the inclined
completeness line that does affect only this bin. 
These bright clusters are as concentrated as the red clusters and do not
have the shallow distribution of the blue ones: a K-S test results in
a 1\% probability that the bright ones are not distributed than the red ones 
and a 95\% probability that they have a different distribution than the blue 
ones.

\section{The Galaxy Light Profile}

It is now of interest to compare the surface density profile of the GCS with
that of the galaxy light and to search for any color gradient, which would
indicate a correspondence between the GCS and the galaxy population itself.

\begin{figure}[t]
\centerline{\resizebox{\hsize}{!}{\includegraphics{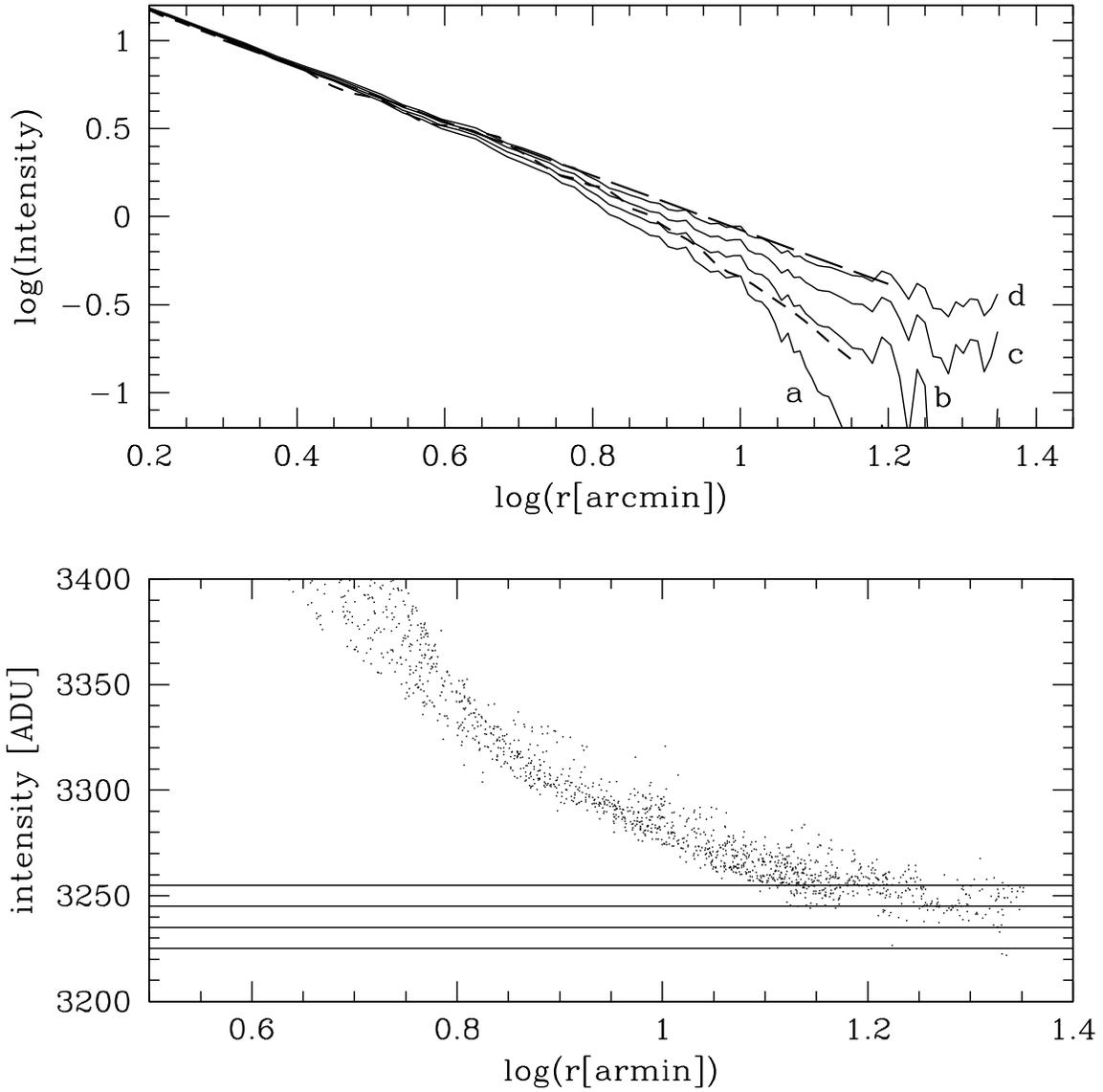}}}
\caption{
	The luminosity profile from our R images (solid lines, for four different
        background values), the one published by Schombert et al. (\cite{schombert86},
        short dashed line) and the one by Caon et al. (\cite{caon94}, long dashed line).
        We shifted all profiles to the B surface magnitude given by Caon et al.
        within the radial range of 1$\arcmin$ to 1.5$\arcmin$.
        The lower panel shows the corresponding background values for the 4 profiles.
	\label{fig:lumprof1}}
\end{figure}

Previous photographic work covering a similarly large field has been carried
out by Schombert (\cite{schombert86}; V-band) and Caon et al.(\cite{caon94};
B-band). The resulting profiles are plotted in Fig.\,\ref{fig:lumprof1}. Beyond
a radius of 5$\arcmin$ they begin to diverge in the sense that the V profile
gets brighter than the B profile. Besides the fact that this ''color gradient''
has the very unexpected sense of getting redder, it also reaches the
unplausibly high value of almost 2 mag at 14$\arcmin$. We conclude that either
V or B, or both, are not reliable at faint surface brightness values. We
therefore tried to derive the light profile from our images as well.

We measured the brightness around all point sources within an annulus between
8\arcsec \ and 11\arcsec \ on the point source-subtracted C and R images. We
defined this brightness to be the mode of the distribution of pixel intensities
after a 5\% clipping of the brightest peaks. Without the $\sigma$
clipping the scatter would have been significantly larger.

Care has to be taken due to the use of eight different chips and corresponding
differences in the sensitivities that are left after flatfielding. To account
for these differences we shifted the zero points of individual chips to a
common mean value by adjusting the sky brightness in adjacent fields. The
resulting correction factors are of the order of 0.3\% of the background level
in the R image and in the order of 1\% in the C images. Since the R images are
deeper we used them to trace the galaxy light to large radii, whereas the C
data has been used to search for a color gradient at smaller radii.  We
excluded regions with bright stars, other galaxies and obvious flatfield
structure (chip $\#4$, chip $\#5$ and upper part of chip $\#8$).  The inner
2$\arcmin$ are saturated.

The most difficult part in determining the galaxy luminosity profile is the
correct determination of the sky background: a tiny 0.5\% deviation in the sky
level has already a profound effect on the galaxy light distribution as shown
in Fig.\,\ref{fig:lumprof1} for the T1 profile.  In the upper panel the
luminosity profile is given for different sky values; these sky levels are
indicated in a linear intensity scale in the lower panel, where the x-axis is
enlarged. Individual measurements are plotted as dots. If we would have reached
the true sky background, profile\,{\bf b} would be appropriate and our
resulting luminosity profile would look nearly identical to that of Caon
et\,al. (\cite{caon94}). However, the measurements suggest that the galaxy
light has not yet leveled out. On the other hand profile {\bf a} can be
excluded, since the observed background is clearly lower that the one belonging
to this profile. The profile by Schombert et al. can be reproduced (profile
{\bf d}) when using the lowest background in the lower panel of
Fig.\,\ref{fig:lumprof1}, but such a low background is almost certainly
excluded since in this case the galaxy would have a radial extension of approximately
500 kpc and hence it would be one of the largest galaxies known.  As will be
discussed later on, profile {\bf c}, based on our prefered baclground also 
resembles well that of the GCs and thus we
adopt it as our preferred profile. The difference with the B-profile of Caon et
al. would still suggest a color which becomes redder with increasing distance,
contrary to what we find (see below). Only profile {\bf a} delivers a color
gradient in the required sense, but this is not supported by our data.

From Fig.\,\ref{fig:lumprof1} it is also apparent that the profile becomes
rather diffuse for radii smaller than approximately 6\arcmin. The reason for
this is the non-negligible ellipticity of NGC 1399. The lower profile envelope
belongs to the minor axis while the upper one belongs to the major axis. We 
determined an average decline only, but this poses no problem for the comparison
with the cluster distribution, which has also been determined assuming a
circular geometry. Moreover, the slope of a power-law is not altered by
averaging over the ellipticity. We nevertheless used this information to derive
the radial dependence of the ellipticity which is shown in the next section.

The C-profile reaches the background level already at 13\arcmin \ due to the
shallower C-data and we determine the background by averaging the ``profile''
at larger radii.

Between $2.5\arcmin$ and $9\arcmin$ (where the effects of the background
uncertainty are still tolerable) we fitted power-laws to the profile and
obtained exponents of $-1.85\pm0.03$ for the T1 profile and of $-1.21\pm0.02$
for the C profile, thus finding a clearly shallower C-profile. This holds for
any sky background which deviates no more than 2\% from our measured
background, to which we assign an error of less than 1\%.  The color gradient
in C-T1 is shown in the lower panel of Fig.\,\ref{fig:surflum_1404_1399} in
Sec.7.2. The color gradient is also plotted as a smoothed curve, together with
error limits that are estimated from the background uncertainties, in
Fig.\,\ref{fig:spec_freq_model}.

We see that the color gradient might level out to a constant value of
C-T1 around 1.2 at 6\arcmin, which is approximately the same radius beyond
which the blue and red GCs start to show indistinguishable surface density
profiles. This might be evidence that the behavior of the GCS is indeed
reflected in the field star population. However, the mean color of the clusters
at this radius is (linearly averaged) C-T1=1.47, and hence considerably redder
than the galaxy color.

\subsection{Covering the whole Radial Range}

\begin{figure}[t]
\centerline{\resizebox{\hsize}{!}{\includegraphics{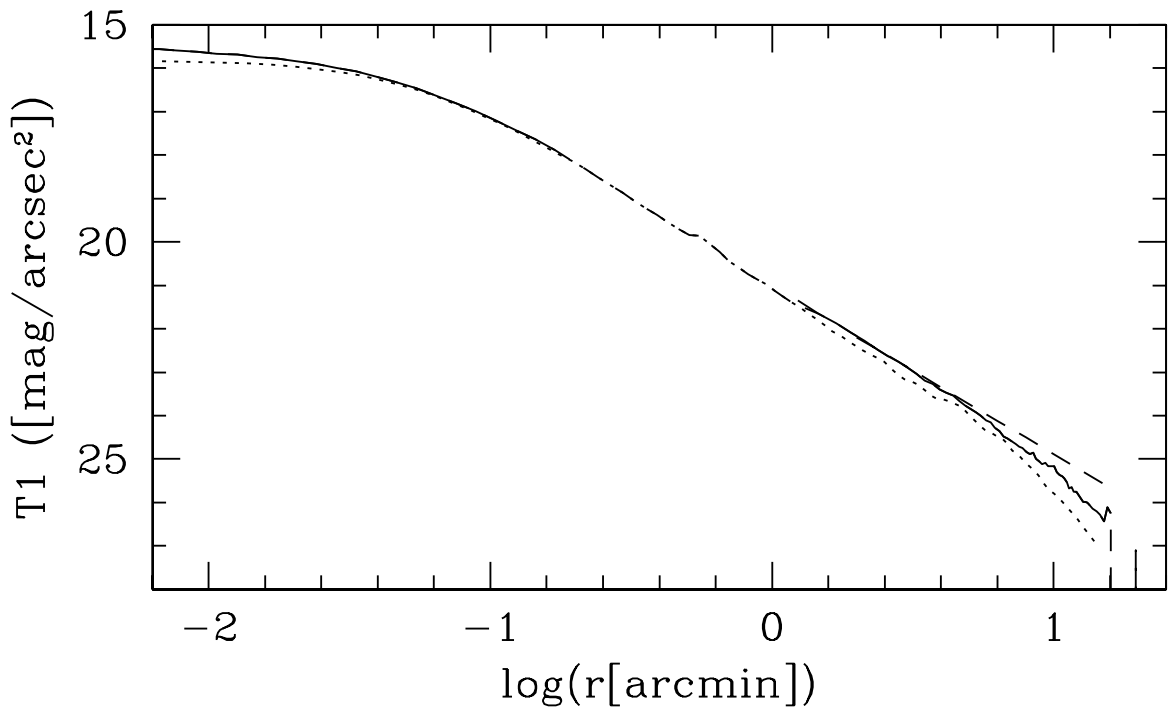}}}
\caption{
	The T1-luminosity profile of NGC 1399. We shifted all profiles
        (even if measured in a different passband) to our profile, which
        is the solid line at large radii. The profile of
        Caon et al. (\cite{caon94}) is the short dashed curve, the one of
        Schombert et al. (\cite{schombert86}) the long dashed curve and
        the HST profile by Lauer et al. (\cite{lauer95}) is the solid line
        at small radii. \label{fig:lumcompare}}
\end{figure}

We are unable to determine the total magnitude of NGC 1399 because of the
missing inner 2\arcmin. However, we can proceed by taking this data from the
literature.  In Fig.~\ref{fig:lumcompare} we plot the available luminosity
profiles which were shifted onto our T1-magnitude scale using our T1-profile.  
We shifted Caon et al.'s and Schombert et al.'s data in the range between
1$\arcmin$ and 1.5$\arcmin$. In the innermost region a V-profile based on HST
observations has been published by Lauer et al.  \cite{lauer95}, which we used
out to 12\arcsec \ . In the range between 12\arcsec \ and 1.5\arcmin, we used
the B profile of Caon et al. \cite{caon94}. For larger radii, we used our
profile. Color gradients in B-R and V-R in the inner region are very small and
should not introduce significant errors (e.g. Idiart et al. \cite{idiart02}).

From this data we can construct a luminosity profile for NGC 1399 that covers 3
orders of magnitude in radius. In Tab.\,\ref{tab:profile} we list this profile,
together with the integrated T1 magnitude at selected radii.

A good fit to the surface brightness profile of NGC 1399 is: 

\begin{equation}\nonumber
T1(r) = 2.125\log(1+(r/0.055\arcmin)^2)+15.75.
\end{equation} 

\begin{figure}[t]
\centerline{\resizebox{\hsize}{!}{\includegraphics{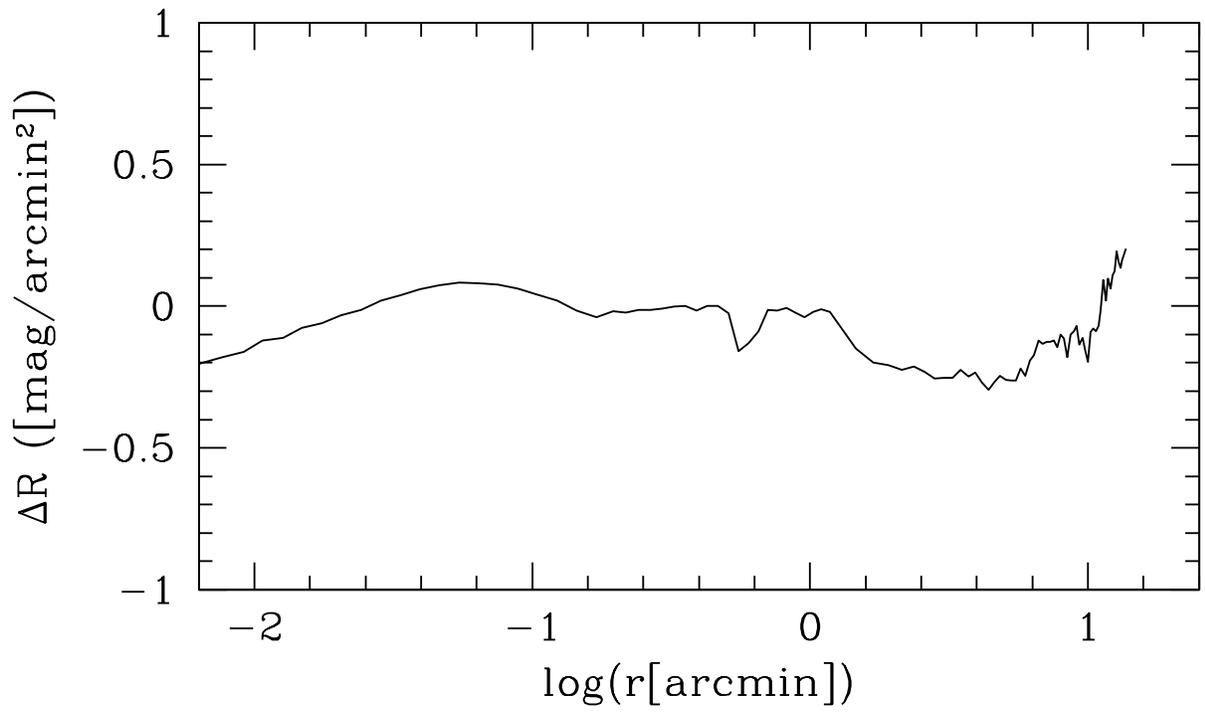}}}
\caption{
	Difference between the observed and fitted galaxy light profile. Positive values mean
        fit values brighter than observed. \label{fig:surf_lum_residuals}}
\end{figure}

The deviation from the fit is shown in Fig.\,\ref{fig:surf_lum_residuals},
where we plot the difference between observed and fitted magnitude. Similarly
to the globular cluster density profile, the fit to the light profile
using the whole radial range also results in a profile being too steep at large
radii (see Fig.\,\ref{fig:surf_lum_residuals}). For these large radii, an
exponent of 1.83 as derived above is more adequate.
Integrating this profile, one obtains the total R-luminosity of NGC 1399:
M$_\mathrm{R}=-23.33$ (assuming a distance modulus of 31.4 and an absolute
solar luminosity of M$_\mathrm{R_{\sun}} =4.28$.).

If we adopt spherical geometry, one can derive the deprojected profile. Because
we have the profile as an analytic expression, an easier way than to apply the
deprojection formula is to adopt a power-law of the above form, use the
projection formula and vary the parameters until the best fit is found.

Using this procedure, the best fit we obtain is:
\begin{equation}\nonumber
 L(T1) [L_{\sun}/pc^3] = 101\cdot (1+r/221\mathrm{pc})^{-2.85}.
\end{equation} 

\begin{figure}[t]
\centerline{\resizebox{\hsize}{!}{\includegraphics{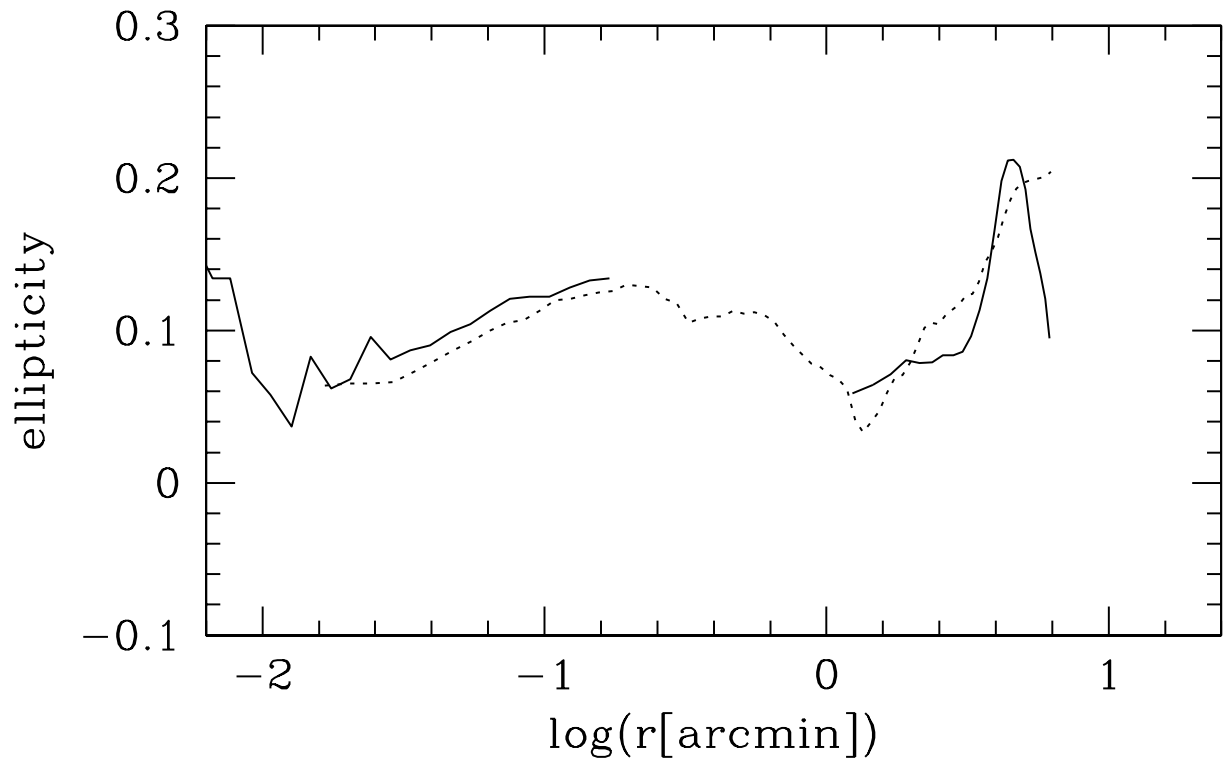}}}
\caption{
	The radial dependence of the ellipticity of the galaxy T1 light
        is plotted as solid lines in the outer region using our MOSAIC data
        (using a fixed position angle at 90$\degr$) and in the inner part
        using that of Lauer et al. (\cite{lauer95}). The dashed line shows the
        ellipticity derived by Caon et al. (\cite{caon94}). \label{fig:ellip}}
\end{figure}

We also determined the ellipticity of NGC 1399. However, we did not try to fit
a position angle to the rather noisy data, but estimated the position angle
rather by eye inspection of the image. We used a
fixed position angle of 90$\degr$. The position angle variation from 110\degr \
to 85\degr \ seen by Caon et al. (\cite{caon94})  is not critical since it occurs
between 30\arcsec \ and 1.35\arcmin, while our profile starts at 2\arcmin. The
resulting radial dependence of the ellipticity is shown in
Fig.\,\ref{fig:ellip} together with Caon et al.'s values.
The two measurements are in good agreement. The ellipticity of NGC\,1399
is 0.1 between $2\arcmin$ and $4\arcmin$, then it rises to about 0.2 at
$5\arcmin$. The further decline at larger radii is rather uncertain, since at
these large radii our azimuthal coverage is not homogeneous due to the
exclusion of areas with bright stars, other galaxies and obviously remaining
flatfield structure.

\begin{figure}[t]
\centerline{\resizebox{\hsize}{!}{\includegraphics{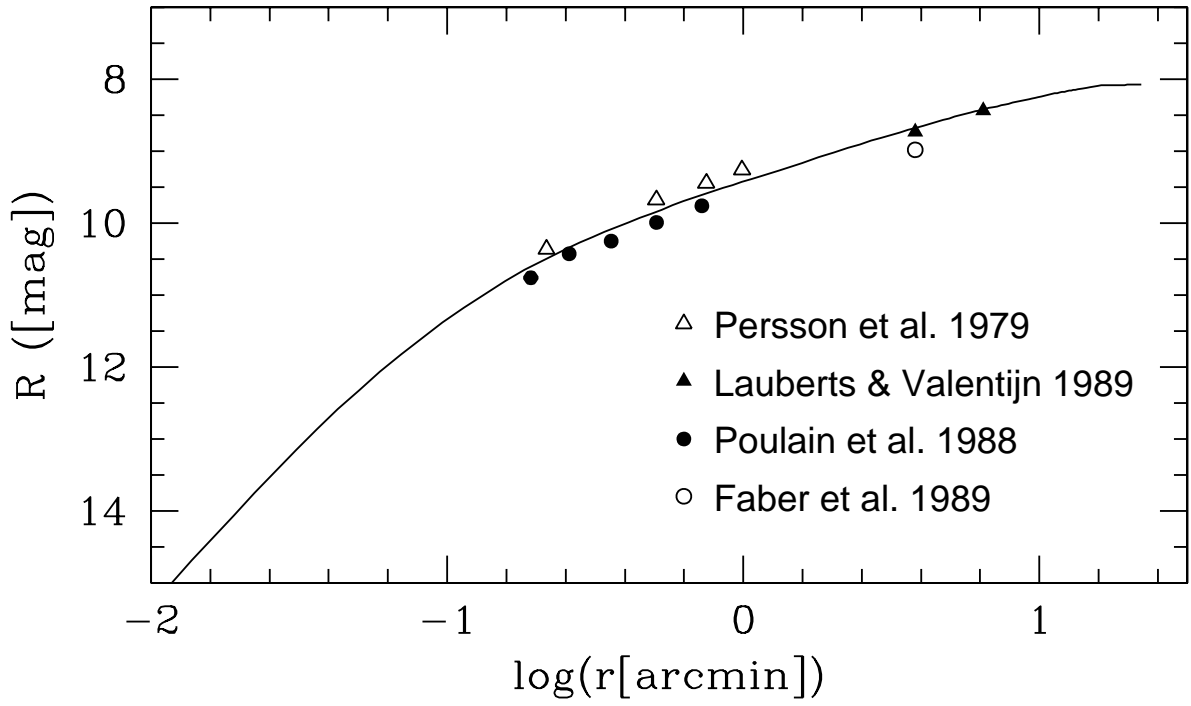}}}
\caption{
	Integrated R magnitude dependence on radius for our data
        and from various literature sources. For this purpos the T1 measurements
        have been transformed into R measurements. \label{fig:gal_lumcomp}}
\end{figure}

In Fig.\ref{fig:gal_lumcomp} we compare the integrated light profile with aperture 
measurements from the literature. This figure shows that our determination
agrees well with the literature values, and that our data do not suggest any
systematic zero point difference.

\section{The Specific Frequency}

The specific frequency is the number of globular clusters per unit luminosity
scaled to a galaxy with an absolute luminosity of $M_V=-15$\,mag ($S_V=N\cdot
10^{0.4(M_V-15)}$). 

In Fig.\,\ref{fig:rad_distr_MOS_VLT_HST}, the galaxy light profile is compared
with the cluster density profile over nearly the whole observed radial range.
Determining the specific frequency means integrating over both curves. In the
inner part the galaxy light is more concentrated than the globular clusters,
which reflects the fact that the core radius of the GCS is about 18 times
larger than that of the field population. This also means that the specific
frequency is not radially constant. We shall proceed in deriving first a global
value, and then study the radial dependence of the specific frequency for both
red and blue clusters.

\subsection{The Global Specific Frequency}

To derive the total number of clusters we have to account for the
incompleteness with respect to the globular cluster luminosity function. In
Sect.\,2.4 we determined this to be $(35\pm5)\%$.

Within the radial range from 2\arcmin \ to 17\arcmin \ (excluding a radius of
$2'$ around NGC\,1404), we find $2600\pm50$ cluster candidates.  Within the
central 2\arcmin \ Forbes et al. (\cite{forbes98}) found 700 clusters. From our
background field we know that a cluster candidate sample brighter than
$\mathrm{T1}=23.5$ has a contamination of $0.48\pm0.02$\,objects/arcmin$^2$.
Accounting for this contamination and incompleteness, we find a total number of
$6100 \pm 770$ clusters within 83\,kpc (15\arcmin).

Alternatively, we integrated the density profile derived in Sec.\,7 and
obtained a number of $6800 \pm 950$ clusters within 83\,kpc. Within 10\arcmin,
this results in 5793 clusters. Averaging the results of both determinations we
find the total number of clusters within 15\arcmin \ to be $6450\pm700$.
Our obtained number agrees well with what has been found for the cluster content within
10\arcmin: Ostrov et al. (\cite{ostrov98}), Forbes et al. (\cite{forbes98}),
and Kissler-Patig et al. (\cite{kisslerpatig97}) all quote $\approx 5700$
clusters within 10\arcmin.  In the inner part we have an excellent agreement
with older studies, too: within 1.5' and 2.5' we determine a total number of
$734\pm114$ clusters, while Ostrov et al. found $746\pm80$, and within 2.5' and
4' we have $921\pm140$ whereas Ostrov et al. give $967\pm100$. 

With this total number of clusters, we find a global specific frequency of
$\mathrm{S}_\mathrm{R}= 3.2 \pm 0.7 $ using a total galaxy magnitude of
T1$=8.08$ (Table 3) and m-M=$31.4\pm0.2$. 
Since we cannot measure the galaxy light in the V band, we
have to convert T1 into V magnitudes to be able to compare our value with the
literature. Using a color of V-R$=0.55$ (Mackie
\cite{mackie90}) we find a specific frequency of $\mathrm{S}_\mathrm{V}= 5.1
\pm 1.2$ within 15\arcmin.

Forbes et al. (\cite{forbes98}) quoted a specific frequency of $11.5 \pm 1$.
They used an absolute magnitude of M$_\mathrm{V}=-21.74$ (taken from Faber et
al. \cite{faber89}), while we would use T1=8.25, which translates to
M$_\mathrm{V}=-22.59$ (see above) for the total luminosity (within 10\arcmin).
Part of the difference comes from the different distance modulus adopted by
Forbes et al.  (m-M=31.2). However, the major discrepancy is in the apparent
magnitude: Faber et al. give the B$_\mathrm{T}$, magnitude which is defined as
the magnitude within the 25\,mag/arcsec$^2$ isophote. Using a B-R=1.8 value, we
find that 25\,mag/arcsec$^2$ corresponds to a radius of 3.8\arcmin. Within this
radius we measure an approximate B luminosity (again assuming no B-R color
gradient) of 10.4 mag, which is in reasonable agreement, considering all
uncertainties to the Faber et al. value of B=10.55.  The same argument can also
be applied to explain the high specific frequencies determined by Kissler-Patig
et al. (\cite{kisslerpatig97}), Wagner et al. (\cite{wagner91}), and Bridges et
al. (\cite{bridges91}). This shows that one of the most important caveats in
deriving the specific frequency is that the clusters and the galaxy light has
to be measured within the same area, a fact that has already been pointed out
by Ostrov et al. (\cite{ostrov98}) for the case of NGC\,1399.

\subsection{The Local Specific Frequency}

Because of the existence of subpopulations in the GCS, it is useful 
to define a local specific frequency where the number of GCs is normalized to
the brightness of the area in which they are counted.

\begin{figure}[t]
\centerline{\resizebox{\hsize}{!}{\includegraphics{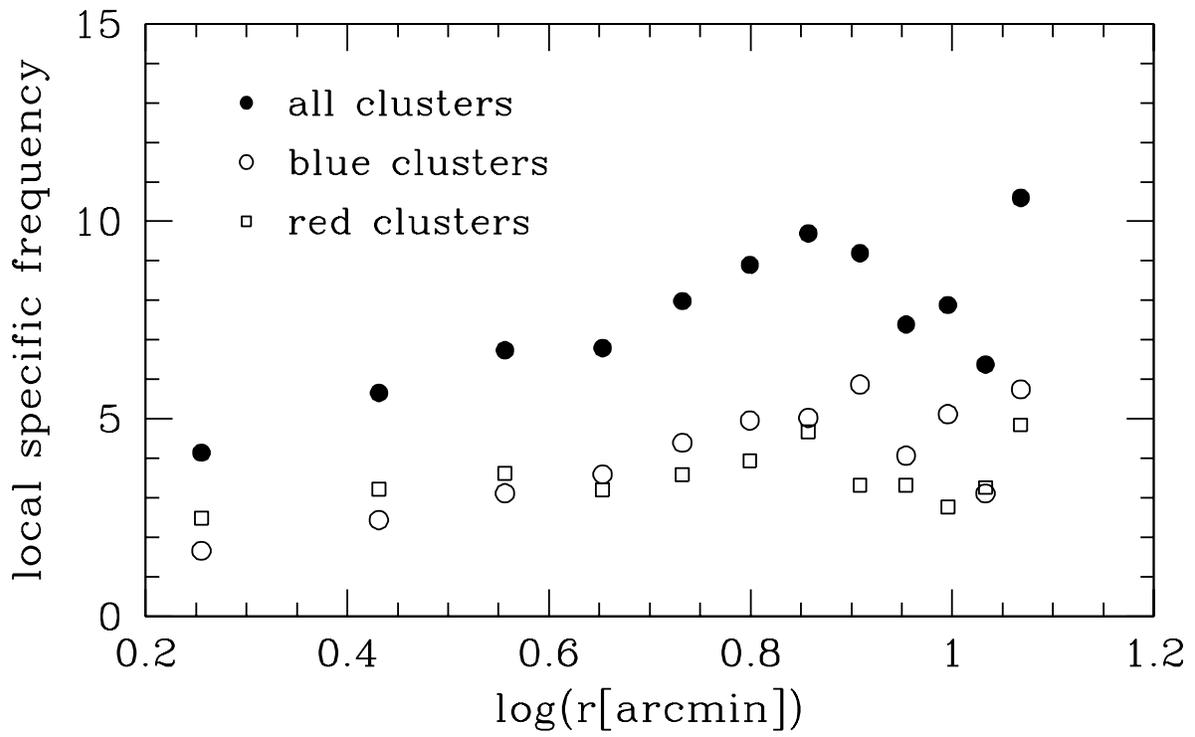}}}
\caption{
	The radial local specific frequency is plotted for the whole cluster
        sample and for the metal-rich and metal-poor subsamples. \label{fig:spec_freq}}
\end{figure}

\begin{figure}[t]
\centerline{\resizebox{\hsize}{!}{\includegraphics{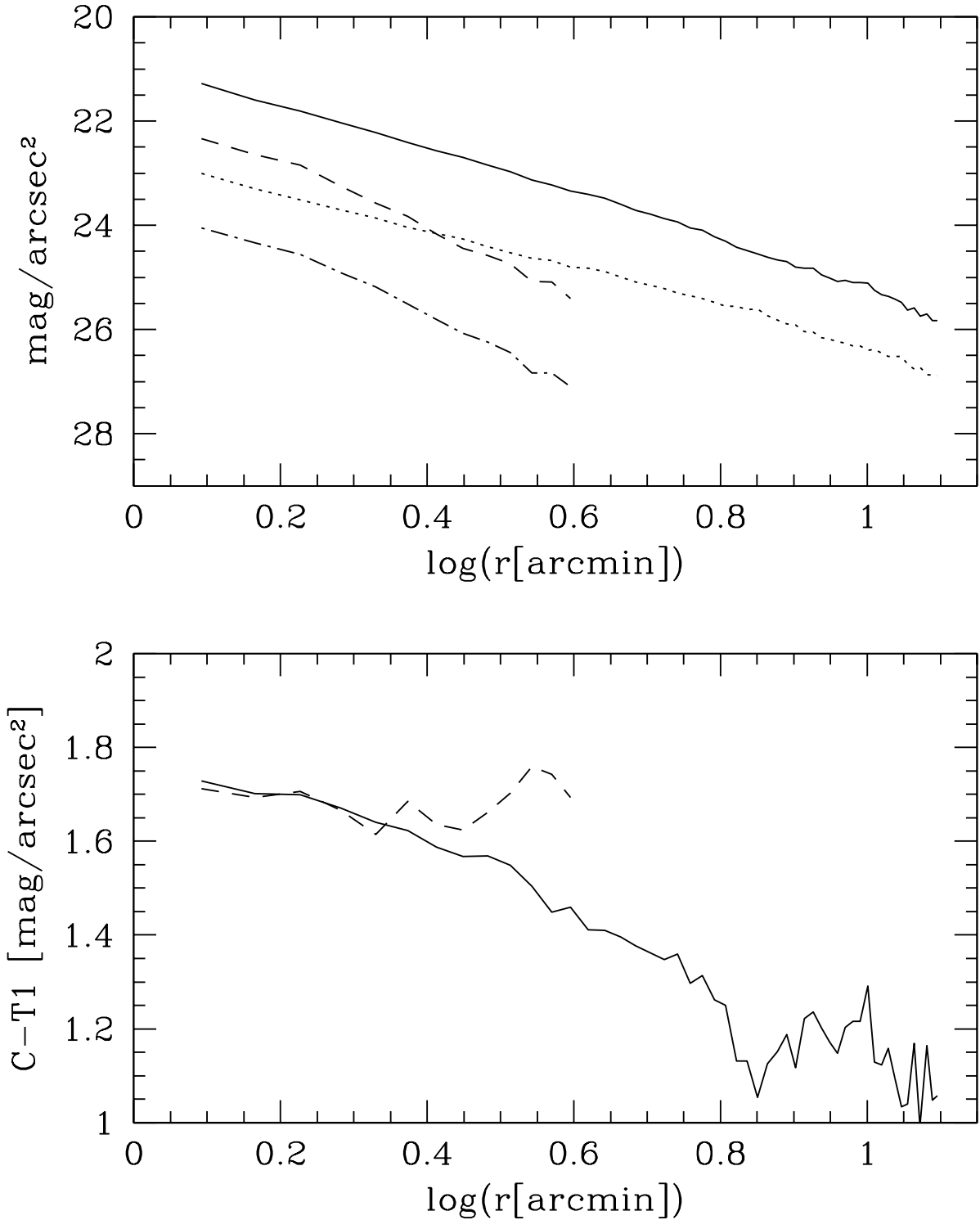}}}
\caption{
	{\bf Upper panel:} The luminosity profile in T1 (solid line) and
        C (dotted line) of NGC\,1399 is be compared to the T1 (dashed line)
        and C (dash-dotted line) surface luminosity of NGC 1404.
        {\bf Lower panel:} The radial color is plotted for
        NGC\,1399 (solid line) and NGC\,1404 (dashed line). \label{fig:surflum_1404_1399}}
\end{figure}

The local specific frequency of the red, the blue and the total cluster sample
for different radial bins is given in Tab.4 and displayed in
Fig.\ref{fig:spec_freq}. Since the galaxy light profile in the $R$ band matches
well the radial distribution of the red cluster population within $11\arcmin$,
its local specific frequency remains nearly constant around S$_R=3$. Outside of
$11\arcmin$ small background uncertainties make a definite statement on the
surface brightness profile difficult. However, using a reasonable background
it is easy to find a galaxy profile that follows the red clusters also at these
large radii. The blue cluster system, on the other hand, shows within
$11\arcmin$ a slightly shallower profile and therefore its specific frequency
with respect to the R band luminosity increases outwards proportional to
$r^{0.8\pm0.2}$. However, the behavior at larger radii is uncertain as well.

\section{Discussion}

\subsection{The Cluster Color Distribution}

In general, there is little doubt that the color of a GC indicates its metallicity in
an old elliptical galaxy like NGC\,1399 (however, we cannot exclude that younger
clusters are mixed in; Forbes et al. \cite{forbes01b}).
The GC color distribution has been the starting point for many investigations
regarding the substructure of GCSs of elliptical galaxies. It has been found by many
authors (e.g. Kundu \& Whitmore \cite{kundu01a}, Forbes\,\&\,Forte \cite{forbes01},
Larsen et al. \cite{larsen01}, Gebhardt\,\&\,Kissler-Patig \cite{gebhardt99}, see
also the corresponding section in the book of Ashman and Zepf \cite{ashman98}) that
GCSs often exhibit two peaks in the color distribution, labeled as ''bimodality''.  
If  bimodality
is not obvious then frequently at least the existence of two peaks
is statistically better supported than an unimodal color distribution.
Usually, a linear transformation between color and metallicity without accounting
to an intrinsic scatter around the mean color metallicity is applied (e.g. Harris
et al. \cite{harris00}, Kissler-Patig et al. \cite{kisslerpatig97}, Harris et al.
\cite{harris91}). This means that a bimodal color distribution translates into a bimodal
metallicity distribution which is assigned to two
different populations. Because bimodal color distributions of clusters
in ellipticals  have been
predicted by Ashman \& Zepf \cite{ashman92} in the context of a simple merger
scenario, color bimodality has been interpreted supporting the formation of 
GCSs of elliptical
host galaxies in merger events. However, despite the fact that a merger {\it can}
produce a bimodal color distribution it is not clear whether a bimodal distribution
{\it necessarily} implies a merger scenario, e.g. even in a monolithic collapse scenario
a bimodal metallicity distribution (of stars) can be obtained: 
Samland et al. (\cite{samland97}) simulated the formation
of a disk galaxy starting with a coherent gas cloud. They found that disk, halo and bulge stars
have different metallicity distributions (their Fig.7). If one plots their 
Fig.7 in logarithmic metallicity bins a bi- or even
multimodal metallicity distribution becomes apparent. They note ''that a bimodal
star formation or a pre-enrichment of the protogalactic cloud is not required by our
model. The metallicity distributions for the different galactic regions result purely from the
local star formation history, the large-scale gas flows, and the phase transitions by
condensation and evaporation.''

We also would like to call attention to the fact that the color-metallicity relation is
non-linear, flattening out in the metal-poor regime (Harris\,\&\,Harris \cite{harris02}). In 
conjunction with an intrinsic spread in the color-metallicity relation due to second 
parameters and photometric erros, this leads to the circumstance that, starting from the bluest 
colors equidistant color intervals are projected onto progressively larger metallicity
intervals, which decrease once the linear regime is reached. What this effect means
for the interpretation in particular of the blue peak, is certainly worth to investigate,
but beyond the scope of our paper.

The question whether the color distribution of a GCS depends on the host
galaxy's properties is still unclear.
Regarding the Washington system only few GCSs have been studied:
Geisler et al. (\cite{geisler96b}) investigated the GCS of NGC 4472 and 
obtained $\mathrm{C-T1}\approx 1.3$ and $\mathrm{C-T1}\approx 1.8$ for the 
colors of the blue and read peak, respectively. The blue peak of NGC\,1427, 
a low luminosity elliptical galaxy, also agrees with the blue peak of the 
NGC\,1399 clusters ($\mathrm{C-T1}\approx1.4$, Forte et al. \cite{forte01}). 
M\,87 also has a GCS with a bimodal color distribution that peaks at 
C-T1$\approx 1.25$, while the position of the red peak remains uncertain
in the color histogram shown by  C\^ot\'e et al. \cite{cote01}, which is based 
on still unpublished photometry. Two more GCSs for which Washington photometry 
have been obtained are those of NGC 3311 (Secker et al. \cite{secker95}) and 
NGC 3923 (Zepf et al. \cite{zepf95}). Both galaxies (which have been observed 
during the same run) show abnormally red color distributions. However, Brodie 
et al. \cite{brodie00} performed V-I photometry with HST in NGC 3311
and found a  normal bimodal distribution with peaks at the expected colors.
Since the Washington run was not photometric, one might conjecture that both
NGC 3311 and NGC 3923 were subject to calibration uncertainties.
Together with our results on NGC 1399
we conclude that at least the blue peaks have
very similar colors in those GCSs of elliptical galaxies investigated 
so far with Washington photometry.

\subsection{Interpreting the Radial Cluster Distribution}

In Paper {\sc ii}, we will show that the blue and red clusters within 7\arcmin \ are also kinematically distinct:
The red clusters have a velocity dispersion of $274\pm15$\,km/s, while the blue clusters have
$310\pm10$\,km/s. This larger dispersion can well be accounted for in the framework of a
spherically dynamical model due to the different surface density slopes. In other
words, a higher/lower dispersion is dynamically equivalent to a shallower/steeper
surface density profile in an isotropic equilibrium situation.

A distinction of two subpopulations on the basis of the color histogram alone bears a high
degree of uncertainty.
However, in light of the
correlation between metallicity and density profile/kinematics, we find it justified
to talk about two populations out to a radius of about $7'$. However, that does not
necessarily imply the existence of two {\it distinct} populations, in the sense of 
different formation epochs or formation mechanisms. It may well be that there is a
continuum of properties, over which our color sampling is simply taking an average.
Going to larger galactocentric radii, the distinction between subpopulations becomes
much less pronounced. The density profiles of blue and red clusters are more or less
the same and the color distribution indicates a broad range of metallicities, even
if two Gaussians are still a better representation that one Gaussian, but as we
said above, we do not assign too much physical significance to that.  

Despite the large field used we have not yet reached the background.
The total extension of the GCS of NGC 1399 is hence not yet known. 
However, for GCs with a very large radii the question remains whether such  clusters
should be considered as belonging to NGC 1399 or as an "intergalactic GC population" 
(White \cite{white87b}, West et al. \cite{west95}). Bassino et al. (\cite{bassino02})
used Washington observations around dwarf elliptical galaxies in the Fornax clusters to 
search for GCs at large radii. They found a GC candidate surface density of
$0.25\pm0.08/\mathrm{arcmin}^2$ and $0.12\pm0.06/\mathrm{arcmin}^2$ at a radial distance of 
40\arcmin \ and 2\degr \ 
from NGC 1399, respectively. For comparison, using their limiting magnitude
of R=22\,mag, we find a surface density of $0.3\pm0.1/\mathrm{arcmin}^2$ at a 
radial distance of 
16\arcmin. Extrapolating our derived radial profile we would expect a 
surface density of $\approx0.14/\mathrm{arcmin}^2$ at $40\arcmin$ and of 
$\approx0.02/\mathrm{arcmin}^2$ at 2\degr. In particular 
at the larger radius we would expect a smaller number of GCs. 
This could be an indication
that a population of "intergalactic clusters" exists, as proposed by Bassino et al.
(\cite{bassino02}).

\subsection{Comparison of the NGC 1399 Light Profile with that of a ''Normal'' Elliptical}

The luminosity profiles of many ''normal'' ellipticals follow a
$r^{1/4}$-de\,Vaucouleurs law. Here we have the opportunity to differentially compare
NGC 1399 with such a ''normal'' galaxy, NGC 1404.  This difference is illustrated in
Fig.\,\ref{fig:surflum_1404_1399}, where the profiles of NGC 1399 and of NGC 1404 are
compared. We derived the luminosity profile of NGC\,1404 in C and T1 in the same way
as for NGC\,1399. Care has been taken to take account of the extended halo of NGC
1399, which envelopes NGC 1404 fully in projection. We subtracted its contribution, 
i.e. the final
luminosity profile, from each pixel. For the sky background the same value as for NGC
1399 has been used.

It can be seen that the light profile of NGC\,1404, in contrast to NGC\,1399, does
not exhibit a uniform power law, but has an exponent of about -2 for radii smaller
than 1.8\arcmin \ and -2.9 for larger radii, i.e. almost -4 in deprojection.

The deviation from a $r^{1/4}$-de\,Vaucouleurs law is not unusual for galaxies
brighter than $M_V = -21$. The light profile shapes of brighter galaxies become
progressively more concave when plotted against $r^{1/4}$ (Kormendy \& Djorgovski
\cite{kormendy89}). These authors define a cD galaxy by a light profile, which
exhibits an inflection point independent from the way it is plotted. In this sense,
NGC 1399 would not be a cD galaxy.

Whatever the exact definition of a cD galaxy may be, it is useful for further insight
to think about the physical reason of a de\,Vaucouleurs profile. Many ''normal``
elliptical galaxies can be well described by a Jaffe \cite{jaffe87} profile, which in
projection resembles the de\,Vaucouleurs law, and which in space in the outer regions
is characterized by a transition from a $r^{-3}$ to a $r^{-4}$ power law. Jaffe
(\cite{jaffe87}) and White (\cite{white87}) have shown that a $r^{-4}$ power law
emerges if the energy distribution function has a sharp break at the escape energy,
given a nearly Keplerian potential.

NGC 1399 has a massive dark halo resembling the potential of an isothermal sphere
out to a distance of at least 40 kpc, shown by the dynamics of the globular clusters 
(Paper {\sc ii}).  The temperature of the X-ray gas outside this
radius remains isothermal (Jones et al. \cite{jones97}). Therefore, even if the exact shape
of the potential is not known, we can at least conclude that the Keplerian regime is
not reached out to the very faint surface brightnesses where our photometry becomes
unreliable. Thus it is very tempting to relate the persistence of the $r^{-3}$
profile to the dark halo of NGC 1399.

We therefore understand the ''halo'' of NGC 1399, which morphologically appears as a
deviation from a de\,Vaucouleurs profile, not as an additional component but as the
natural continuation of the galaxy profile in the isothermal potential of the dark
matter halo. Thus the term ''cD halo'' might be misleading since it implies an
additional stellar component.

Finally, we do not detect a color gradient for NGC\,1404. This shows that the
population composition of NGC\,1404 is radially more homogeneous than that of
NGC\,1399. Additionally, this also demonstrates that the observed color gradient 
in NGC 1399 is not an artefact of an incorrect background subtraction.

\subsection{Disentangling the Stellar Populations}

Superior to the approach to relate the two observed red and blue cluster subpopulations to the
integral galaxy light would be the comparison between clusters and the respective
field populations that formed during the same star formation event. This is not
possible in a direct way. However, we can follow a more indirect approach in which we model
the galaxy C and R light profile using some simplifying assumptions.  We assume the
field population to be a superposition of only two populations, one corresponding to
the blue clusters and one to the red clusters. These two field populations shall have
the mean color of their corresponding cluster populations. The relation between the
number of clusters and the luminosity of their corresponding field population will be
called {\it intrinsic specific frequency} (ISF) and is measured with respect to the
T1 filter.

\begin{figure}[t]
\centerline{\resizebox{\hsize}{!}{\includegraphics{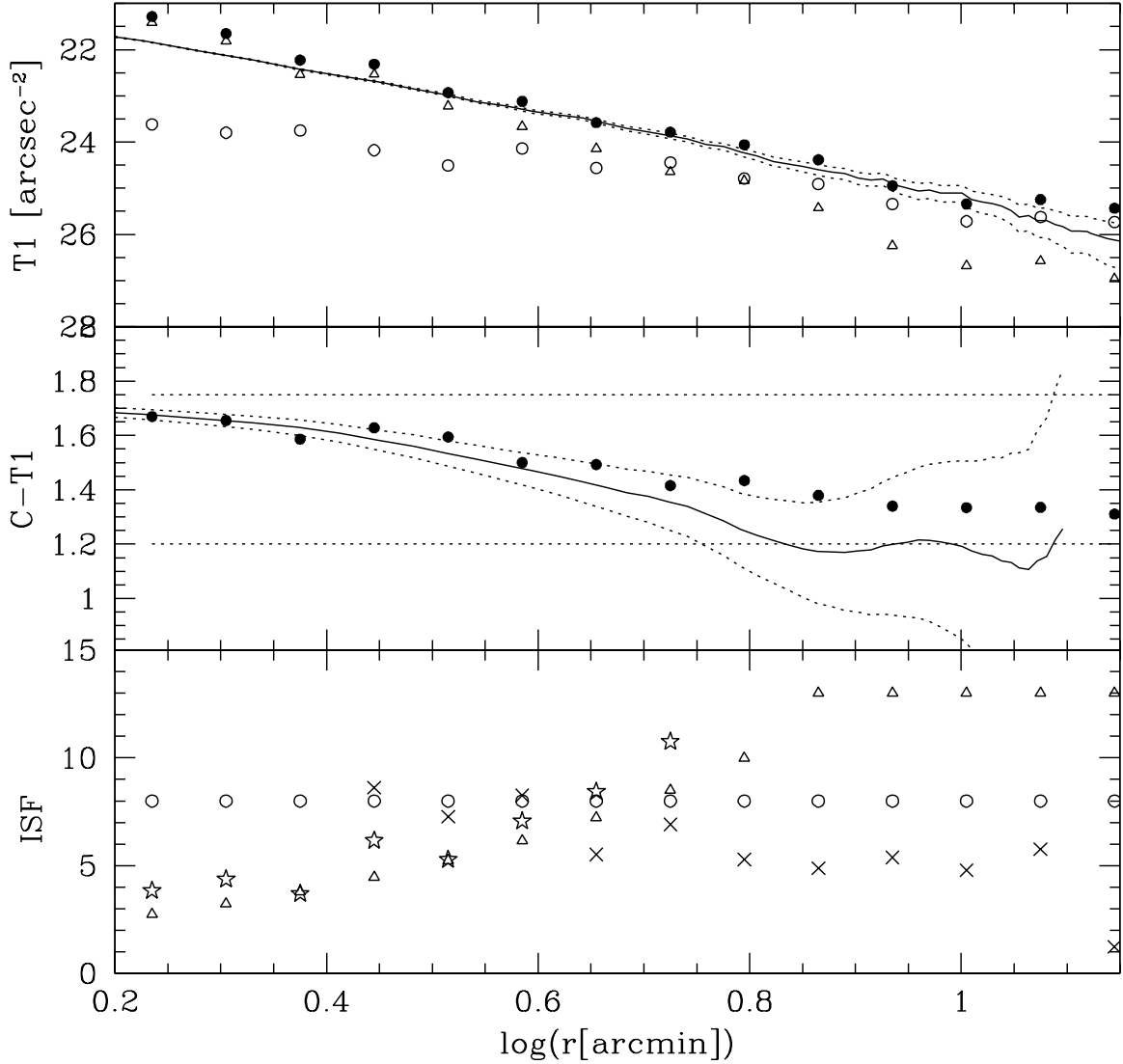}}}
\caption{
	A simple model to fit the galaxy light profile measured in the C and R
        band. In the {\bf upper panel} the galaxy light is shown as solid lines
        accompanied by estimated errors due to background variations. The solid
        circles show the model light distribution, that is a composite of the
        contribution of the red (triangles) and blue (open circles) populations.
        The two horizontal lines indicate the color that has been used as mean color
        for the two populations.
        In the {\bf middle panel} the color profile is plotted with a solid line 
        together with the estimated errors and compared to the model color distribution
        (solid circles). In the {\bf lowest panel} we plotted as open circle and
        open triangle the two ISF that define the model for blue and red population
        respectively. The crosses and the diamonds show the ISF for blue and red
        populations, respectively, when it is directly calculated. \label{fig:spec_freq_model}}
\end{figure}

It is not possible to fit both, the color and the luminosity profile, if both cluster
populations would have the same constant ISF because of their different surface
density profiles. In addition, to reproduce the color gradient requires to allow for a change of the
properties of the red populations by either a radial color or radial local specific
frequency dependence. We limit ourselves to consider a changing ISF for the red
population, which might be the strongest assumption we make.
Otherwise the problem would not be  well constrained and many
solutions could be found. We used the linear approximation $\mathrm{ISF}\propto r$. A
shallower variation, e.g. $\mathrm{ISF}\propto \log (r)$ results in a too shallow
color gradient. With a much steeper dependence, e.g. $\mathrm{ISF} \propto r^2$ the
model luminosity profile shows a considerable dip at the point where the light of the
blue population begins to dominate the total light, which is also incompatible with
the observations. A model that fits the data reasonably well is depicted in
Fig.\,\ref{fig:spec_freq_model}. A characteristic of any satisfactory model is that
the ISF of the blue cluster population is much higher than that of the red one at
smaller radii.

It is also possible to use the color and luminosity profile together with the density
profiles of red and blue clusters to derive the ISF for both clusters samples directly.
However, the disadvantage of this approach is that it is much stronger
 influenced by noise. Nevertheless, it may serve as a consistency test and
we include the results of this calculation in 
Fig.\,\ref{fig:spec_freq_model}. This illustrates that
the two methods agree well considering the uncertainties.

\subsection{From Light Distribution to Mass Distribution}

We note that the color gradient in NGC 1399 indicates a radial change in the stellar
M/L profile. If the outer regions are more metal-poor the M/L value decreases
towards larger radii. If one wants to use the luminosity profile to infer the mass profile, one
has to account for this effect.  For an  estimation, we assume two populations
with the same age of 12 Gyr of different mean metallicities: 
a metallicity of -0.4 dex (representing the
inner part in accordance with the galaxy's color) and -1.4 dex 
(representing the outermost parts). 
Worthey's interpolation
engine then gives for the respective M/L-values in R (Miller-Scalo IMF): 1.45 (-0.4
dex) and 1.23 (-1.4 dex). This means that the associated mass profile becomes somewhat
steeper than what would be derived from the luminosity profile using a constant M/L.
Thus an increase of the power law exponent from $-1.83$ to $\approx-2.2$ is 
expected.

\section{Summary and conclusions}

We investigated  the GCS of NGC 1399 in the bands Washington C and Kron R  
with the CTIO 4-m MOSAIC camera. The imaged area is $36\arcmin\times36\arcmin$,
thus being among the largest concerning  GCS investigations.

The main findings of this study are:

\begin{itemize}

\item
The relatively modest specific frequency of $5.1\pm1.2$ of the GCS of NGC 1399, already noted by 
Ostrov et al. (\cite{ostrov98})
is confirmed. The local specific frequency increases with radius and is
higher for the blue than for the red clusters.

\item The color distribution shows a very pronounced bimodal appearance.
However, we argue that the blue peak might artificially be caused by the non-linear color
metallicity relation in conjunction with the color scatter at a given metallicity.

\item The surface density profile is shallower for the blue clusters than
for the red clusters within 8 arcmin. Outside a radius of 8 \arcmin, 
both surface density profiles are not distinguishable.

\item The galaxy light shows a colour gradient resembling  the change of the
surface density profiles of blue and red clusters. The galaxy light can
be modeled with the simple assumption of only two populations, each traced
by a cluster population of the same mean color. A model consistent
with the data has a constant ISF for the blue clusters and a 
radially increasing ISF for the red clusters.

\item The brightest clusters do not show the pronounced bimodal color
distribution of the fainter ones and have an at least as concentrated radial 
distribution as the red GCs.

\end{itemize}

NGC 1399 was one of the ''classic'' high specific frequency galaxies.
It would be very interesting to perform new studies of the similar cases M87 and NGC 3311
in order to see whether
their  high specific frequency still withstand  closer scrutiny. If not, central galaxies
in galaxy clusters
are perhaps not the places of extraordinarily high efficiency of globular cluster formation, 
for which they have been taken.

The GC color distribution, in particular bimodality alone is not very restrictive
for galaxy formation scenarios. Various models have been published that
explain the observed bimodality/metallicity distribution, e.g. Ashman \& Zepf (\cite{ashman92})
Forbes et al. (\cite{forbes97}),  C\^ot\'e et al. (\cite{cote01}), Beasley et al. (\cite{beasley02}).
It is believed that bimodal color distributions can be used as an
argument against monolithic collapse scenarios, for example Rhode \& Zepf (\cite{rhode01}). 
However, doubts remain:
Samland et al. (\cite{samland97}) showed that also a monolithic collapse scenario result in
populations that have different metallicity distributions. 

Since the color distribution alone is not suitable to constrain formation scenarios it is 
necessary to use further information. 

The close relation of the color gradient
of NGC 1399 with the mean color of the GCS might be
seen in favor of a strongly dissipative period during which the globular
clusters and the stellar body of the galaxy were formed together.
However,
a counterargument is the radially increasing local specific frequency
of the red clusters. During a dissipative collapse one would expect the star
formation rate to be highest where the density is highest, i.e in the central regions.
Accordingly, also the efficiency of globular cluster formation should
decrease radially outwards, if the star formation rate is a determining parameter (Larsen
\& Richtler \cite{larsen00}).

Additionally, one has to take into account the kinematic and dynamical
properties of the GCS. These are (Paper {\sc ii}): blue 
and red clusters
show different velocity dispersions in accordance with their different 
surface density profiles. The orbit distribution is nearly isotropic. Only
the outer blue clusters (if any) show signatures of rotation. The overall mass distribution
resembles an isothermal sphere.

Thus some sort of relaxation mechanism must have been active in the early
times of NGC 1399, which is difficult to find in a purely dissipative
collapse, but naturally provided by a phase of violent relaxation during
a merger. The merger components (probably more than only 2) could have been 
gas-rich, unevolved metal-poor galaxies where cluster formation was efficient 
during the starburst preceding the merger itself. In addition to the 
(presumably mainly metal poor) cluster populations of the merger components, the
bulk of clusters, both metal-rich and metal-poor, could have been formed during this 
period. 
These clusters have after the merging took place roughly the same spatial
distribution.
However,
after the merger remnant has reached its equilibrium, cluster formation continued 
in the inner regions, albeit  with a decreased efficiency, forming the metal
rich part of the inner red cluster population.  

Based on this idea we can estimate the number of additional red 
globular clusters. For this we assume that we have one population of metal rich clusters
that have the same density profile as the blue ones (albeit with a different normalization
taken from the red clusters at radii larger than 8\arcmin). Then we require $\approx 700$
additional red clusters to explain the difference between the two profiles.

\section*{Acknowledgments}
We acknowledge the help of M.Shara, D.Zurek and E.Grebel, who obtained some
MOSAIC images that were used in this investigation. BD gratefully acknowledges
financial support of the Alexander-von-Humboldt Foundation via a Feodor Lynen
Stipendium. DG and WPG gratefully acknowledge support from the Chilean Center
for Astrophysics FONDAP No. 15010003. We thank an anonymous referee for 
helpful comments. 

\section*{References}
\begin{quote}

\bibitem[1987]{arimoto87} Arimoto, N., Yoshii, Y., 1987,
	A\&A 173, 23
\bibitem[1992]{ashman92} Ashman, K.M., Zepf, S.E., 1992,
	ApJ 384, 50
\bibitem[1993]{ashman93} Ashman, K.M., Zepf, S.E., 1993,
	MNRAS 264, 611 
\bibitem[1995]{ashman95} Ashman, K.M., Conti, A., Zepf, S.E., 1995,
	AJ 110, 1164
\bibitem[1998]{ashman98} Ashman, K.M., Zepf, S.E., 1998,
	''Globular Cluster Systems'', Cambridge Astrophysics Series,
	Cambridge University Press 
\bibitem[2002]{bassino02} Bassino L.P., Cellone S.A., Forte J.C., Dirsch B., 2002,
	A\&A submitted
\bibitem[2002]{beasley02} Beasley, M.A., Baugh, C.M., Ducan, A.F. et al., 2002,
	MNRAS 333, 383
\bibitem[2002]{bekki02} Bekki, K., Forbes, D.A., Beasley, M.A., Couch, W.J., 2002,
	astro-ph/0206008, MNRAS in press
\bibitem[1996]{bertin96} Bertin, E., Arnouts, S., 1996,
	A\&AS 117, 393
\bibitem[1991]{bridges91} Bridges, T.J., Hanes, D.A., Harris, W.E., 1991,
	AJ 101, 469
\bibitem[2000]{brodie00} Brodie J.P., Larsen S.S., Kissler-Patig M., 2000,
	ApJ 543, L19
\bibitem[2000]{buote00} Buote, D.A., 2000, 
	MNRAS 311, 176
\bibitem[1994]{caon94} Caon, N., Capaccioli, M., D'Onofrio, M., 1994,
	A\&AS 106, 199
\bibitem[2001]{cote01} C\^ot\'e, P., Mc\,Laughlin, D.E., Hanes, D.A., et al., 2001
	ApJ 559, 828
\bibitem[2002]{cote02} C\^ot\'e, P., West, M.J., Marzke, R.O., 2002,
	ApJ 567, 853
\bibitem[1976]{dawe76} Dawe, J.A., Dickens, R.J., 1976,
	Nature 263, 395
\bibitem[1999]{elmegreen99} Elmegreen, B.G., 1999,
	Ap\&SS 269, 469   
\bibitem[1989]{faber89} Faber, S.M., Wegner, G., Burstein, D. et al., 1989
	ApJS 69, 763
\bibitem[1997]{forbes97} Forbes, D.A., Brodie, J.P., Grillmair, C.J., 1997,
	AJ 113, 1652
\bibitem[1998]{forbes98} Forbes, D.A., Grillmair, C.J., Williger, G.M. et al., 1998,
	MNRAS 293, 325
\bibitem[2001a]{forbes01} Forbes, D.A., Forte, J.C., 2001,
	MNRAS 322, 257
\bibitem[2001b]{forbes01b} Forbes, D.A., Beasley, M.A., Brodie, J.P., Kissler-Patig M., 2001,
	ApJ 563, L143
\bibitem[2001]{forte01} Forte, J.C., Geisler, D., Ostrov, P.G. et al., 2001,
	AJ 121, 1992
\bibitem[1999]{gebhardt99} Gebhardt, K., Kissler-Patig, M., 1999,
	AJ 118, 1526
\bibitem[1996]{geisler96a} Geisler, D., 1996,
	AJ 111, 480
\bibitem[1996]{geisler96b} Geisler, D., Lee, M.G., Kim, E., 1996,
	AJ 111, 1529
\bibitem[1994]{grillmair94} Grillmair, C.J., Freeman, K.C., Bicknell, G.V., et al., 1994,
	ApJ 422, L9
\bibitem[1999]{grillmair99} Grillmair, C.J., Forbes, D.A., Brodie, J.P., Elson, R.A.W., 1999,
	AJ 117, 167
\bibitem[1986]{hanes86} Hanes, D.A., Harris, W.E., 1986,
	ApJ 309, 564
\bibitem[1977]{harris77} Harris, H.C., Canterna, R., 1977,
	AJ 82, 798
\bibitem[1981]{harris81} Harris, W.E., van den Bergh, S., 1981,
	ApJ 309, 654
\bibitem[1987]{harris87} Harris, W.E., Hanes, D.A., 1987,
	AJ 93, 1368 
\bibitem[1991]{harris91} Harris, W.E., Allwright, J.W.B., Pritchet, C.J., van den Bergh, S., 1991,
	ApJS 76, 115
\bibitem[1996]{harris96} Harris, W.E., Hanes, D.A., 1987,
	CDS VizieR On-line Data Catalog: VII/195
\bibitem[2000]{harris00} Harris, W.E., Kavelaars, J.J., Hanes, D.A. et al., 2000,
	ApJ 533, 137
\bibitem[2001]{harris01} Harris, W.E., 2001, in ''Star Clusters``, 
	Saas-Fee Advanced Course 28, Lecture Notes 1998, eds. L. Labhardt and B. Binggeli, 
        Springer-Verlag, Berlin, p.223
\bibitem[2002]{harris02} Harris W.E., Harris G.L.H., 2002,
	AJ 123, 3108
\bibitem[2000]{hilker99} Hilker, M., Infante, L., Richtler, T., 1999,
	A\&AS 138, 55 
\bibitem[2002]{idiart02} Idiart, T.P., Michard, R., de Freitas Pacheco, J.A., 2002,
	A\&A 383, 30
\bibitem[1987]{jaffe87} Jaffe, W., 1987,
	in ''Structure and Dynamics of Elliptical Galaxies'', IAU Symp. 127, ed. T. de Zeeuw,
	D. Reidel, Dordrecht, p.511
\bibitem[1997]{jones97} Jones, C., Stern, C., Forman, W., 1997,
	ApJ 482, 143
\bibitem[1988]{killeen88} Killeen, N.E.B., Bicknell, G.V., 1988,
	ApJ 325, 165	
\bibitem[1994]{kissler94} Kissler, M., Richtler, T., Held, E. et al., 1994, 
	A\&A 287, 463
\bibitem[1997]{kisslerpatig97} Kissler-Patig, M., Kohle, S., Hilker, M. et al., 1997,
	A\&A 319, 470
\bibitem[1989]{kormendy89} Kormendy, J., Djorgovski, S., 1989,
	ARAA 27, 235
\bibitem[2001a]{kundu01a} Kundu, A., Whitmore, B.C., 2001a,
	AJ 121, 2950 
\bibitem[2001b]{kundu01b} Kundu, A., Whitmore, B.C., 2001b,
	AJ 122, 1251	
\bibitem[1999]{larsen99} Larsen, S.S., Richtler, R., 1999, 
	A\&A 345, 59
\bibitem[2000]{larsen00} Larsen, S.S., Richtler, R., 2000, 
	A\&A 354, 836
\bibitem[2001]{larsen01} Larsen, S.S., Brodie, J.P., Huchra, J.P. et al., 2001,
	AJ 121, 2974
\bibitem[1989]{lauberts89} Lauberts, A., Valentijn, E.A., 1989,
	''The surface photometry catalogue of the ESO-Uppsala galaxies'',
	European Southern Observatory, Garching	
\bibitem[1995]{lauer95} Lauer, T.R., Ajhar, E.A., Byun, Y. et al., 1995,
	AJ 110,  2622
\bibitem[2002]{liu02} Liu, M.C., Graham, J.R., Charlot, S., 2002,
	ApJ 564, 216
\bibitem[1999]{mclaughlin99} McLaughlin, D.E., 1999,
	AJ 117, 2398
\bibitem[1990]{mackie90} Mackie, G., Visvanathan, N., Carter, D., 1990,
	ApJS 73, 637
\bibitem[1994]{merritt94} Merritt, D., Tremblay, B., 1994,
	AJ 111, 2243
\bibitem[1993]{ostrov93} Ostrov, P.G., Geisler, D., Forte, J.C., 1993,
	AJ 105, 1762   
\bibitem[1998]{ostrov98} Ostrov, P.G., Forte, J.C., Geisler, D., 1998,
	AJ 116, 2854
\bibitem[1997]{pagel97} Pagel B.E.J., 1997,
	Nucleosynthesis and Chemical Evolution of Galaxies, Cambridge University Press
\bibitem[1979]{persson79} Persson, S.E., Frogel, J.A., Aaronson, M., 1979,
	ApJS 39, 61
\bibitem[1988]{poulain88} Poulain, P., 1988,
	A\&AS 72, 215
\bibitem[2000]{richtler00} Richtler, T., Drenkhahn, G., G\'omez, M., Seggewiss, W.,
         2000, in ''From Extrasolar Planets to Cosmology: The VLT Opening
         Symposium'', eds. J. Bergeron and A. Renzini, Springer Verlag, Berlin,
         p. 259 
\bibitem[2001]{richtler01} Richtler, T., Dirsch, B., Geisler, D. et al., 2001,
	IAU Symp. 207, ''Extragalactic Star Clusters'', eds. Geisler D., Grebel E.K., Minniti D., p. 263
\bibitem[2002]{richtler02} Richtler, T., Dirsch, B., Gebhardt, K. et al., 2002, 
	(Paper {\sc ii}), submitted to AJ 
\bibitem[2001]{rhode01} Rhode, K.L., Zepf, S.E., 2001,
	AJ 121, 210
\bibitem[1997]{samland97} Samland, M., Hensler, G., Theis, Ch., 1997,
	ApJ 476, 544
\bibitem[1998]{schlegel98} Schlegel, D., Finkbeiner, D.,  Davis, M., 1998,
  	ApJ 500, 525.
\bibitem[1986]{schombert86} Schombert, J.M., 1986,
	ApJS 60, 603
\bibitem[1995]{secker95} Secker J., Geisler D., Mc\,Laughlin D.E., Harris W.E., 1995,
	AJ 109, 1019
\bibitem[2001]{tonry01} Tonry, J.L., Dressler, A.D., Blakeslee, J.P. et al., 2001,
	AJ 546, 681
\bibitem[1983]{valentjin83} Valentijn, E.A., 1983,
       	A\&A 118, 123 
\bibitem[2000]{vandenbergh00} van den Bergh, S., 2000,	
	PASP 112, 932 
\bibitem[1991]{wagner91} Wagner, S., Richtler, T., Hopp, U., 1991,
	A\&A 241, 399
\bibitem[1995]{west95} West M.J., C\^ot\'e P. Jones C. et al., 1995,
	AJ 453, L77
\bibitem[1987]{white87b} White III R.E., 1987,
	MNRAS 227, 185
\bibitem[1987]{white87} White, S.D.M., 1987,
	in ''Structure and Dynamics of Elliptical Galaxies'', IAU Symp. 127, ed. T. de Zeeuw,
	D. Reidel, Dordrecht, p.263
\bibitem[1995]{whitmore95} Whitmore, B.C., Schweizer, F., 1995,
	AJ 109, 960
\bibitem[1994]{worthey94} Worthey, G., 1994,
	ApJS 95, 107
\bibitem[1993]{zepf93} Zepf, S.E., Ashman, K.M., 1993,
	MNRAS 264, 611
\bibitem[1995]{zepf95} Zepf, S.E., Ashman, K.M., Geisler D., 1995,
	ApJ 443, 570

\end{quote}

\begin{deluxetable}{ccccccc}
\tabletypesize{\scriptsize}
\tablecaption{Observational data used in this investigation.} 
\tablewidth{0pt}
\tablehead{
	\colhead{Telescope} & 
	\colhead{Date} & 
	\colhead{right ascension} & 
	\colhead{declination} & 
	\colhead{Filter} & 
	\colhead{Exposure time} & 
	\colhead{Seeing}
}
  \label{tab:obssummary}
\startdata
MOSAIC  & 7.12.1999 & 3\degr 38\arcmin 12\arcsec.0& $-35^\mathrm{h}32^\mathrm{m}25^\mathrm{s}$ & R & 420\,sec & $1.1\arcsec$\\
MOSAIC  & 7.12.1999 & 3\degr 38\arcmin 12\arcsec.0& $-35^\mathrm{h}32^\mathrm{m}25^\mathrm{s}$ & R & 420\,sec & $1.0\arcsec$\\
MOSAIC  & 7.12.1999 & 3\degr 38\arcmin 12\arcsec.0& $-35^\mathrm{h}32^\mathrm{m}25^\mathrm{s}$ & R & 420\,sec & $0.9\arcsec$\\
MOSAIC  & 11.12.1999 & 3\degr 38\arcmin 12\arcsec.0& $-35^\mathrm{h}32^\mathrm{m}25^\mathrm{s}$ & R & 420\,sec & $1.2\arcsec$\\
MOSAIC  & 11.12.1999 & 3\degr 38\arcmin 12\arcsec.0& $-35^\mathrm{h}32^\mathrm{m}25^\mathrm{s}$ & R & 420\,sec & $1.2\arcsec$\\
MOSAIC  & 11.12.1999 & 3\degr 38\arcmin 12\arcsec.0& $-35^\mathrm{h}32^\mathrm{m}25^\mathrm{s}$ & R & 420\,sec & $1.3\arcsec$\\
MOSAIC  & 12.12.1999 & 3\degr 38\arcmin 12\arcsec.0& $-35^\mathrm{h}32^\mathrm{m}25^\mathrm{s}$ & R & 420\,sec & $1.2\arcsec$\\
MOSAIC  & 15.12.1999 & 3\degr 38\arcmin 12\arcsec.0& $-35^\mathrm{h}32^\mathrm{m}26^\mathrm{s}$ & C & 1500\,sec & $1.1\arcsec$\\
MOSAIC  & 15.12.1999 & 3\degr 38\arcmin 13\arcsec.0& $-35^\mathrm{h}32^\mathrm{m}15^\mathrm{s}$ & C & 1500\,sec & $1.1\arcsec$\\
MOSAIC  & 15.12.1999 & 3\degr 38\arcmin 13\arcsec.0& $-35^\mathrm{h}32^\mathrm{m}32^\mathrm{s}$ & C & 1500\,sec & $1.0\arcsec$\\
VLT	& 1.12.2000 & 3\degr38\arcmin45\arcsec & $-35^\mathrm{h}23^\mathrm{m}30^\mathrm{s}$ & V & 300\,sec & $0.55\arcsec$\\
VLT	& 1.12.2000 & 3\degr38\arcmin45\arcsec & $-35^\mathrm{h}23^\mathrm{m}30^\mathrm{s}$ & I & 300\,sec & $0.55\arcsec$\\
VLT	& 2.12.2000 & 3\degr38\arcmin08\arcsec & $-35^\mathrm{h}31^\mathrm{m}00^\mathrm{s}$ & V & 300\,sec & $0.64\arcsec$\\
VLT	& 2.12.2000 & 3\degr38\arcmin08\arcsec & $-35^\mathrm{h}31^\mathrm{m}00^\mathrm{s}$ & I & 300\,sec & $0.6\arcsec$\\
VLT	& 3.12.2000 & 3\degr37\arcmin30\arcsec & $-35^\mathrm{h}31^\mathrm{m}30^\mathrm{s}$ & V & 300\,sec & $0.56\arcsec$\\
VLT	& 3.12.2000 & 3\degr37\arcmin30\arcsec & $-35^\mathrm{h}31^\mathrm{m}30^\mathrm{s}$ & I & 300\,sec & $0.5\arcsec$ \\
MOSAIC  & 31.12.1999 & 3\degr 48\arcmin 00\arcsec.0& $-35^\mathrm{h}35^\mathrm{m}00^\mathrm{s}$ & R & 120\,sec & $1.0\arcsec$\\
MOSAIC  & 31.12.1999 & 3\degr 48\arcmin 00\arcsec.0& $-35^\mathrm{h}35^\mathrm{m}00^\mathrm{s}$ & R & 600\,sec & $1.0\arcsec$\\
MOSAIC  & 31.12.1999 & 3\degr 48\arcmin 00\arcsec.0& $-35^\mathrm{h}35^\mathrm{m}00^\mathrm{s}$ & C & 1800\,sec & $1.0\arcsec$\\
\enddata
\end{deluxetable} 

\begin{deluxetable}{cccccc}
\tabletypesize{\scriptsize}
\tablecaption{Color distribution of clusters for the whole sample (N$_\mathrm{all}$),
	within $1.8\arcmin<r<4.5\arcmin$ (N$_1$), within $4.5\arcmin<r<13.5\arcmin$ 
	(N$_2$) and $r>13.5\arcmin$ (N$_3$).}
\tablewidth{0pt}
\tablehead{
	\colhead{C-R} &
	\colhead{N$_\mathrm{all}$} &
	\colhead{N$_1$}&
	\colhead{N$_2$}&
	\colhead{N$_3$}&
	\colhead{N$_\mathrm{BG}$}
}
\startdata
0.8 & $14\pm 13 $ & $2\pm2$   & $10 \pm 7   $& $4 \pm 8   $ & $80 \pm 9$ \\
0.9 & $60\pm 18 $ & $6 \pm 3 $ & $31 \pm 9  $ & $26 \pm 11$  & $129 \pm 11$ \\
  1 & $53\pm 18 $ & $10 \pm 4$ & $41 \pm 11 $ & $3 \pm 11 $  & $135 \pm 12$ \\
1.1 & $169\pm 20$ & $31 \pm 6$ & $96 \pm 13 $ & $42 \pm 12$  & $119 \pm 11$ \\
1.2 & $384\pm 24$ & $59 \pm 8$ & $203 \pm 16$ & $119 \pm 15$ & $108 \pm 10$ \\
1.3 & $458\pm 25$ & $73 \pm 9$ & $253 \pm 17$ & $123 \pm 14$ & $87 \pm 9$ \\
1.4 & $360\pm 22$ & $72 \pm 9$ & $187 \pm 15$ & $88 \pm 12 $ & $53 \pm 7$ \\
1.5 & $223\pm 17$ & $50 \pm 7$ & $109 \pm 11$ & $50 \pm 9  $ & $42 \pm 6$ \\
1.6 & $236\pm 17$ & $55 \pm 7$ & $111 \pm 11$ & $55 \pm 9  $ & $23 \pm 5$ \\
1.7 & $186\pm 15$ & $59 \pm 8$ & $83 \pm 10 $ & $37 \pm 7  $ & $19 \pm 4$ \\
1.8 & $219\pm 16$ & $72 \pm 8$ & $95 \pm 10 $ & $41 \pm 7  $ & $15 \pm 4$ \\
1.9 & $156\pm 13$ & $50 \pm 7$ & $68 \pm 9  $ & $24 \pm 6  $ & $10 \pm 3$ \\
  2 & $108\pm 12$ & $37 \pm 6$ & $45 \pm 7  $ & $15 \pm 5  $ & $13 \pm 4$ \\
2.1 & $39\pm 8  $ & $9 \pm 3 $ & $18 \pm 5  $ & $4 \pm 3   $ & $9 \pm 3$ \\
2.2 & $14\pm 6  $ & $7 \pm 2 $ & $1 \pm 3   $ & $4 \pm 4   $ & $12 \pm 3$ \\
2.3 & $5\pm 4   $ & $1 \pm 1 $ & $3 \pm 2   $ & $1 \pm 1   $ & $5 \pm 2$ \\
\enddata
\label{tab:colordistr}
\end{deluxetable}

\begin{deluxetable}{ccc}
\tabletypesize{\scriptsize}
\tablecaption{Observed radial number density of red ({\it n(red)})  and
	blue ({\it n(blue)}) clusters.} 
\tablewidth{0pt}
\tablehead{
	\colhead{log(r[arcmin])} & 
	\colhead{log(n(red)[arcmin$^{-2}$])} & 
	\colhead{log(n(blue)[arcmin$^{-2}$])}
}
\startdata
0.24 & $ 1.13 \pm 0.06 $ & $0.61 \pm 0.10$  \\
0.31 & $ 1.04 \pm 0.05 $ & $0.54 \pm 0.09$ \\
0.38 & $ 0.82 \pm 0.06 $ & $0.56 \pm 0.08$ \\
0.45 & $ 0.89\pm 0.05  $ & $0.39 \pm 0.09$ \\
0.52 & $ 0.69 \pm 0.06 $ & $0.26 \pm 0.10$  \\
0.59 & $ 0.58 \pm 0.06 $ & $0.40 \pm 0.07$ \\
0.66 & $ 0.46 \pm 0.05 $ & $0.24 \pm 0.07$ \\
0.73 & $ 0.33 \pm 0.05 $ & $0.28 \pm 0.06$ \\
0.80 & $ 0.32 \pm 0.05 $ & $0.14 \pm 0.06$ \\
0.87 & $ 0.20 \pm 0.05 $ & $0.10 \pm 0.05$ \\
0.94 & $ -0.13 \pm 0.06 $ & $-0.08 \pm 0.06$ \\
1.01 & $ -0.30 \pm 0.06 $ & $-0.23 \pm 0.06$ \\
1.08 & $ -0.26 \pm 0.06 $ & $-0.19 \pm 0.05$ \\
1.15 & $ -0.42 \pm 0.06 $ & $-0.24 \pm 0.05$ \\
1.22 & $ -0.43 \pm 0.06 $ & $-0.57 \pm 0.08$ \\
1.29 & $ -0.45 \pm 0.06 $ & $-0.72 \pm 0.10$ \\
1.36 & $ -0.75 \pm 0.10 $ & $-0.88 \pm 0.13$  \\
\enddata
\label{tab:raddistr}
\end{deluxetable}

\begin{deluxetable}{ccccc}
\tabletypesize{\scriptsize}
\tablewidth{0pt}
\label{tab:profile}
\tablecaption{The final luminosity profile of NGC 1399 plotted in Fig.\ref{fig:lumprof1}.
	The second column gives the surface brightness at the radius given in the 
	first column. The third column gives the ellipticity and the fourth column 
	the integrated magnitudes at selected radii. The last column shows the 
	C-T1-color.}
\tablehead{
	\colhead{log(r[arcmin])} &
	\colhead{T1 [arcsec$^{-2}$]} &
	\colhead{ellipticity} & 
	\colhead{T1 [mag]} &
	\colhead{C-T1}
}
\startdata
      -3.436 & 15.39 & 0 & ...& ...\\
      -2.436 & 15.48 & 0.13 & ...& ...\\
      -2.115 & 15.59 & 0.13 & ...& ...\\
      -1.757 & 15.78 & 0.062 & ...& ...\\
      -1.405 & 16.19 & 0.085 & ...& ...\\
      -1.052 & 16.99 & 0.12 & ...& ...\\
	-1.015 & ... & ... & 11.38&\\
     -0.7095 & 18.14 & 0.13 & ...& ...\\
     -0.4893 & 19.05 & 0.11 & ...& ...\\
     -0.2939 & 19.84 & 0.11 & ...& ...\\
     -0.1183 & 20.56 & 0.09 & ...& ...\\
	0.0 & ... & ... & 9.43& ...\\
     0.04123 & 21.27 & 0.07 & ...& 1.71\\
      0.3299 & 22.43 & 0.09 & ...& 1.64\\
      0.5136 & 23.18 & 0.11 & ...& 1.55\\
      0.6422 & 23.69 & 0.20 & ...& 1.41\\
      0.7413 & 24.14 & 0.20 & ...& 1.36\\
	0.75 & ... & ... & 8.48& ...\\
       0.822 & 24.63 & 0.20 & ...& 1.13\\
        0.89 & 24.91 & 0.20 & ...& 1.19\\
      0.9488 & 25.22 & 0.20 & ...& 1.17\\
       1.001 & 25.31 & 0.20 & ...& 1.29\\
       1.047 & 25.69 & 0.20 & ...& 1.03\\
	1.05 & ... & ... & 8.20& ...\\
       1.089 & 26.03 & 0.20 & ...& 1.05\\
       1.123 & 26.24 & 0.20 & ...& ...\\
       1.191 & 26.21 & 0.20 & ...& ...\\
       1.292 & 26.75 & 0.20 & ...& ...\\
	1.292 & ... & ... & 8.08 & ... \\
\enddata
\end{deluxetable}

\begin{deluxetable}{cccc}
\tabletypesize{\scriptsize}
\tablewidth{0pt}
\tablecaption{Local specific frequency for the whole sample (second column), 
	the red subpopulation (third column), and the blue sample (last column).
	The errors include the error of the luminosity profile and Poisson
	counting statistics.}
\tablehead{
	\colhead{log(r[arcmin])}& 
	\colhead{S$_\mathrm{R}$(all)} & 
	\colhead{S$_\mathrm{R}$(red)} &
	\colhead{S$_\mathrm{R}$(blue)}
}
\startdata
  0.26 &$ 4.3 \pm 0.4$ & $2.6 \pm 0.32$ & $1.7 \pm 0.3$\\
  0.43 &$ 5.8 \pm 0.7$ & $3.3 \pm 0.5$ & $2.5 \pm 0.5$\\
  0.56 &$  6.9 \pm 1.0$ & $3.7 \pm 0.7$ & $3.2 \pm 0.7$\\
  0.65 &$ 7.0 \pm  1.2$ & $3.3 \pm 0.8$ & $3.7 \pm 0.9$\\
  0.73 &$ 8.2 \pm   1.5$ & $3.7 \pm  1.0$ & $4.5 \pm  1.1$\\
  0.80 &$ 9.1 \pm  1.9$ & $4.0 \pm  1.3$ & $5.1 \pm   1.4$\\
  0.86 &$ 9.8 \pm  2.3$ & $4.7 \pm  1.6$ & $5.1 \pm  1.6$\\
  0.91 &$ 9.3 \pm   2.4$ & $3.4 \pm  1.5$ & $6.0 \pm  1.9$\\
  0.95 &$ 7.5 \pm  2.4$ & $3.3 \pm  1.6$ & $4.1 \pm    1.8$\\
  1.00 &$ 8.1 \pm  2.7$ & $2.8 \pm  1.6$ & $5.3 \pm  2.2$\\
   1.03 &$ 6.4 \pm  2.7$ & $3.3 \pm  1.9$ & $3.1 \pm  1.9$\\
   1.07 &$ 10.8 \pm  3.9$ & $4.9 \pm  2.7$ & $5.9 \pm  2.9$\\
     1.10 &$ 14.2 \pm  5.0$ & $4.9 \pm  2.9$ & $9.2 \pm  4.0$\\
    1.13 &$ 12.6 \pm  5.0$ & $5.5 \pm  3.3$ & $7.1 \pm  3.7$\\
   1.16 &$ 11.8 \pm   5.4$ & $5.3 \pm  3.6$ & $6.5 \pm  4.0$\\
   1.20 &$  18.3 \pm  9.0$ & $9.9 \pm  6.6$ & $8.4  \pm 6.1$\\\hline
\enddata
\end{deluxetable}

\clearpage

\clearpage

\end{document}